\documentclass[a4paper,12pt,onecolumn,draftclsnofoot]{IEEEtran}

\usepackage{epsfig}
\usepackage{graphicx}
\usepackage{cite}
\usepackage{amsmath}
\usepackage{amssymb}

\newcommand{\comment}[1]  {}
\def\BE{\begin{equation}}
\def\EE{\end{equation}}
\def\BEA{\begin{eqnarray}}
\def\EEA{\end{eqnarray}}

\newcommand{\pdpd}[2]{\frac{\partial ^{2} #1}{\partial #2 ^{2}}}

\DeclareMathOperator*{\argmax}{arg\,max}
\newtheorem{thm}{Theorem}
\newtheorem{lem}[thm]{Lemma}
\newtheorem{alg}{Algorithm}
\newtheorem{prop}[thm]{Proposition}
\newtheorem{defn}[thm]{Definition}
\newtheorem{assm}[thm]{Assumption}

\newtheorem{corol}[thm]{Corollary}
\newtheorem{remark}[thm]{Remark}

\newcommand\ie{{\textsl{i.e.\,}}}
\newcommand\eg{{\textsl{e.g.\,}}}
\newcommand\etal{{\textsl{et al.\,}}}

\newcommand\vb{{\bf b}}
\newcommand\vc{{\bf c}}

\newcommand\vn{{\bf n}}

\newcommand\vr{{\bf r}}

\newcommand\vx{{\bf x}}
\newcommand\vy{{\bf y}}
\newcommand\vz{{\bf z}}
\newcommand\mA{{\bf A}} 
\newcommand\mB{{\bf B}}

\newcommand\mD{{\bf D}}

\newcommand\mI{{\bf I}}

\newcommand\mM{{\bf M}}

\newcommand\mP{{\bf P}}

\newcommand\mR{{\bf R}}
\newcommand\mS{{\bf S}}

\begin{document}

\title{Gaussian Belief Propagation for Solving\\Systems of Linear Equations:\\Theory and Application}

\author{Ori~Shental\thanks{The first two authors contributed equally to this work. }\thanks{O. Shental, P.~H. Siegel and J.~K. Wolf are with the Center for Magnetic Recording Research (CMRR), University of California - San Diego (UCSD), 9500 Gilman Drive, La Jolla, CA 92093, USA (Email: \{oshental,psiegel,jwolf\}@ucsd.edu).}, Danny~Bickson$^{\ast}$\thanks{$^{\ast}$ Contact Author. D. Bickson and D. Dolev are with the School of Computer Science and Engineering, Hebrew University of Jerusalem, Jerusalem 91904, Israel (Email: \{daniel51,dolev\}@cs.huji.ac.il).
        \newline The material in this paper was presented in part at the
        forty-fifth Annual Allerton Conference on Communication, Control, and
        Computing, September 2007 and on the IEEE International Symposium on Information Theory (ISIT), July 2008.
        \newline\textbf{Paper submitted to IEEE Transactions on Information Theory, \today.}}, Paul H. Siegel, Jack K. Wolf and Danny Dolev}
\maketitle

\begin{abstract}
The canonical problem of solving a system of linear equations
arises in numerous contexts in information theory, communication
theory, and related fields. In this contribution, we develop a
solution based upon Gaussian belief propagation (GaBP) that does
not involve direct matrix inversion. The iterative nature of our
approach allows for a distributed message-passing implementation
of the solution algorithm. We address the properties of the GaBP
solver, including convergence, exactness, computational
complexity, message-passing efficiency and its relation to
classical solution methods. We use numerical examples and
applications, like linear detection, to illustrate these
properties through the use of computer simulations. This empirical
study demonstrates the attractiveness (\eg, faster convergence
rate) of the proposed GaBP solver in comparison to conventional
linear-algebraic iterative solution methods.
\end{abstract}

\begin{keywords}
Belief propagation, sum-product, system of linear equations,
probabilistic inference, max-product, Gaussian elimination,
iterative solution methods, Moore-Penrose pseudoinverse, linear
detection, Poisson's equation.
\end{keywords}

\newpage
\section{Introduction}
Solving a system of linear equations \mbox{$\mA\vx=\vb$} is one of
the most fundamental problems in algebra, with countless
applications in the mathematical sciences and engineering.
Given an observation vector \mbox{$\vb\in\mathbb{R}^{m}$} and the data
matrix \mbox{$\mA\in\mathbb{R}^{m\times n}$} ($m\ge
n\in\mathbb{Z}$), a unique solution,
\mbox{$\vx=\vx^{\ast}\in\mathbb{R}^{n}$}, exists if and only if
the data matrix $\mA$ has full column rank. Assuming a nonsingular matrix $\mA$, the system of equations
can be solved either directly or in an iterative manner. Direct
matrix inversion methods, such as Gaussian elimination (LU
factorization,~\cite{BibDB:BookMatrix}-Ch. 3) or band Cholesky
factorization (\cite{BibDB:BookMatrix}-Ch. 4), find the solution
with a finite number of operations, typically, for a dense
$n\times n$ matrix, of the order of $n^{3}$. The former is
particularly effective for systems with unstructured dense data
matrices, while the latter is typically used for structured dense
systems.

Iterative methods~\cite{BibDB:BookAxelsson} are inherently
simpler, requiring only additions and multiplications, and have
the further advantage that they can exploit the sparsity of the
matrix $\mA$ to reduce the computational complexity as well as the
algorithmic storage requirements~\cite{BibDB:BookSaad}. By
comparison, for large, sparse and amorphous data matrices, the
direct methods are impractical due to the need for excessive
matrix reordering operations.

The main drawback of the iterative approaches is that, under
certain conditions, they converge only asymptotically to the exact
solution $\vx^{\ast}$~\cite{BibDB:BookAxelsson}. Thus, there is
the risk that they may converge slowly, or not at all. In
practice, however, it has been found that they often converge to
the exact solution or a good approximation after a relatively
small number of iterations.

A powerful and efficient iterative algorithm, belief propagation
(BP,~\cite{BibDB:BookPearl}), also known as the sum-product
algorithm, has been very successfully used to solve, either
exactly or approximately, inference problems in probabilistic
graphical models~\cite{BibDB:BookJordan}.

In this paper, we reformulate the general problem of solving a
linear system of algebraic equations as a probabilistic inference
problem on a suitably-defined graph. We believe that this is the
first time that an explicit connection between these two
ubiquitous problems has been established\footnote{Recently, we have found out
the work of Moallemi and Van Roy \cite{MinSum} which discusses the connection between
the Min-Sum message passing algorithm and solving quadratic programs.
Both works~\cite{Allerton,MinSum} where published in parallel, and the algorithms where derived independently, using different techniques. In Appendix \ref{MinSum} we discuss the connection between the two algorithms, and show
they are equivalent.}. Furthermore, for the first time,
we provide a full step-by-step derivation of the GaBP algorithm from the
belief propagation algorithm.

As an important consequence, we demonstrate that Gaussian BP (GaBP) provides an
efficient, distributed approach to solving a linear system that
circumvents the potentially complex operation of direct matrix
inversion. Using the seminal work of Weiss and
Freeman~\cite{BibDB:Weiss01Correctness} and some recent related
developments~\cite{BibDB:jmw_walksum_nips,BibDB:mjw_walksum_jmlr06,MinSum,ISIT1,Allerton,ISIT2},
we address the convergence and exactness properties of the
proposed GaBP solver. Other properties of the GaBP solver, as
computational complexity, message-passing efficiency and its
relation to classical solution methods are also investigated.

As application of this new approach to solving a system of linear
equations, we consider the problem of linear detection using a
decorrelator in a code-division multiple-access (CDMA) system.
Through the use of the iterative message-passing formulation, we
implement the decorrelator detector in a distributed manner. This
example allows us to quantitatively compare the new GaBP solver
with the classical iterative solution methods that have been
previously investigated in the context of a linear implementation
of CDMA
demodulation~\cite{grant99iterative,BibDB:TanRasmussen,BibDB:YenerEtAl}.
We show that the GaBP-based decorrelator yields faster convergence
than these conventional methods. Furthermore, the GaBP convergence
is further accelerated by incorporating the linear-algebraic
methods of Aitken and Steffensen~\cite{BibDB:BookHenrici} into the
GaBP-based scheme. As far as we know, this represents the first
time these acceleration methods have been examined within the
framework of message-passing algorithms.

The paper is organized as follows. We introduce the problem model in Section \ref{sec:preliminaries}. In Section~\ref{sec:derivation}, we derive the distributed
GaBP-based solution method and address its convergence and
exactness properties in Section \ref{sec:converge-exact}. We discuss the algorithm
complexity and message passing efficiency in Section \ref{sec:complexity}.
The relation to classical linear algebra solution methods is explored in Section \ref{sec:classical}.
In Section \ref{sec_results}, we outline numerical examples and use the
linear detection problem to illustrate experimentally the superior convergence rate of the GaBP solver,
relative to conventional iterative methods. Concluding remarks are
presented in Section~\ref{sec:conclusion}.

%
%
%

\section{Preliminaries: Notations and Definitions}
\label{sec:preliminaries}
\subsection{Linear Algebra}
We shall use the following linear-algebraic notations and
definitions. The operator $\{\cdot\}^{T}$ stands for a vector or
matrix transpose, the matrix $\mI_{n}$ is an $n\times n$ identity
matrix, while the symbols $\{\cdot\}_{i}$ and $\{\cdot\}_{ij}$
denote entries of a vector and matrix, respectively. Let
$\mM\in\mathbb{R}^{n\times n}$ be a real symmetric square matrix
and $\mA\in\mathbb{R}^{m\times n}$ be a real (possibly
rectangular) matrix. Let

\begin{defn}[Pseudoinverse]
The Moore-Penrose pseudoinverse matrix of the matrix $\mA$,
denoted by $\mA^{\dag}$, is defined as \BE
\mA^{\dag}\triangleq{(\mA^{T}\mA)}^{-1}\mA^{T}. \EE
\end{defn}

\begin{defn}[Spectral radius]
The spectral radius of the matrix $\mM$, denoted by $\rho(\mM)$,
is defined to be the maximum of the absolute values of the
eigenvalues of $\mM$, \ie, \BE \rho(\mM)\triangleq\max_{1\leq
i\leq s}(|\lambda_{i}|),\EE where $\lambda_{1},\ldots\lambda_{s}$
are the eigenvalues of the matrix $\mM$.
\end{defn}

\begin{defn}[Diagonal dominance]
The matrix $\mM$ is\\
\begin{enumerate}
  \item weakly diagonally dominant if \BE |M_{ii}|\ge\sum_{j\neq
  i}|M_{ij}|,\forall i,\EE
  \item strictly diagonally dominant if \BE |M_{ii}|>\sum_{j\neq
  i}|M_{ij}|,\forall i,\EE
  \item irreducibly diagonally dominant if $\mM$ is irreducible\footnote{
  A matrix is said to be reducible if there is a permutation
  matrix $\mP$ such that $\mP\mM\mP^{T}$ is block upper triangular. Otherwise, it is irreducible.},
  and \BE |M_{ii}|\ge\sum_{j\neq
  i}|M_{ij}|,\forall i,\EE with strict inequality for at least one
  $i$.
\end{enumerate}
\end{defn}

\begin{defn}[PSD]
The matrix $\mM$ is positive semi-definite (PSD) if and only if
for all non-zero real vectors $\vz\in\mathbb{R}^{n}$, \BE
    \vz^{T}\mM\vz\geq0.
\EE
\end{defn}

\begin{defn}[Residual]
For a real vector $\vx\in\mathbb{R}^{n}$, the residual,
$\vr=\vr(\vx)\in\mathbb{R}^{m}$, of a linear system is
$\vr=\mA\vx-\vb$.
\end{defn}
The standard norm of the residual, $||\vr||_{p}
(p=1,2,\ldots,\infty)$, is a good measure of the accuracy of a
vector $\vx$ as a solution to the linear system. In our
experimental study, the Frobenius norm (\ie, $p=2$) per equation
is used, $||\vr||_{F}/m=\sqrt{\sum_{i=1}^{m}r_{i}^{2}}/m$.

\begin{defn}
The condition number, $\kappa$, of the matrix $\mM$ is defined as
\BE \kappa_{p}\triangleq||\mM||_{p}||\mM||_{p}^{-1}.\EE For $\mM$
being a normal matrix (\ie, $\mM^{T}\mM=\mM\mM^{T}$), the
condition number is given by \BE
\kappa=\kappa_{2}=\Big|\frac{\lambda_{\textrm{max}}}{\lambda_{\textrm{min}}}\Big|,
\EE where $\lambda_{\textrm{max}}$ and $\lambda_{\textrm{min}}$
are the maximal and minimal eigenvalues of $\mM$, respectively.
\end{defn}

Even though a system is nonsingular it could be ill-conditioned.
Ill-conditioning means that a small perturbation in the data
matrix $\mA$, or the observation vector $\vb$, causes large
perturbations in the solution, $\vx^{\ast}$. This determines the
difficulty of solving the problem. The condition number is a good
measure of the ill-conditioning of the matrix. The better the
conditioning of a matrix the condition number is smaller, going to
unity. The condition number of a non-invertible (singular) matrix
is set arbitrarily to infinity.

\subsection{Graphical Models} We will make use of the following
terminology and notation in the discussion of the GaBP algorithm.
Given the data matrix $\mA$ and the observation vector $\vb$, one
can write explicitly the Gaussian density function, $p(\vx)$, and
its corresponding graph $\mathcal{G}$ consisting of edge
potentials (compatibility functions) $\psi_{ij}$ and self
potentials (`evidence') $\phi_{i}$. These graph potentials are
simply determined according to the following pairwise
factorization of the Gaussian function~(\ref{eq_G})\BE
p(\vx)\propto\prod_{i=1}^{n}\phi_{i}(x_{i})\prod_{\{i,j\}}\psi_{ij}(x_{i},x_{j}),\EE
resulting in \mbox{$\psi_{ij}(x_{i},x_{j})\triangleq
\exp(-x_{i}A_{ij}x_{j})$} and
\mbox{$\phi_{i}(x_{i})\triangleq\exp\big(b_{i}x_{i}-A_{ii}x_{i}^{2}/2\big)$}.
The edges set $\{i,j\}$ includes all non-zero entries of $\mA$ for
which $i>j$. The set of graph nodes $\textrm{N}(i)$ denotes the
set of all the nodes neighboring the $i$th node (excluding node $i$). The set
$\textrm{N}(i)\backslash j$ excludes the node $j$ from
$\textrm{N}(i)$.

\subsection{Problem Formulation}
Let $\mA \in \mathbb{R}^{m\times n}$ ($m,n\in\mathbb{N}^{\ast}$)
be a full column rank, $m \times n$ real-valued matrix, with $m
\ge n$, and let $\vb \in \mathbb{R}^m$ be a real-valued vector.
Our objective is to efficiently find a solution $\vx^{\ast}$ to
the linear system of equations $\mA\vx=\vb$ given by\BE
\label{eq_solution} \vx^{\ast}=\mA^{\dag}\vb. \EE

Throughout the development in this contribution, we make the
following assumption.
\begin{assm}\label{assum_square}
The  matrix $\mA$ is square (\ie, $m=n$) and symmetric.
\end{assm}
For the case of square matrices the pseudoinverse matrix is
nothing but the data matrix inverse, \ie,
\mbox{$\mA^{\dag}=\mA^{-1}$}. For any linear system of equations
with a unique solution, Assumption~\ref{assum_square} conceptually
entails no loss of generality, as can be seen by considering the
invertible system defined by the new symmetric (and PSD) matrix
${\mA^{T}}_{n\times m}\mA_{m\times n}\mapsto\mA_{n\times n}$ and
vector ${\mA^{T}}_{n\times m}\vb_{m\times 1}\mapsto\vb_{n\times
1}$. However, this transformation involves an excessive
computational complexity of $\mathcal{O}(n^{2}m)$ and
$\mathcal{O}(nm)$ operations, respectively. Furthermore, a sparse
data matrix may become dense due to the transformation, an
undesired property as far as complexity concerns. Thus, we first
limit the discussion to the solution of the popular case of square
matrices. In Section~\ref{sec_new_const} the proposed GaBP solver is
extended to the more general case of linear systems with
rectangular $m\times n$ full rank matrices.

\section{The GaBP-Based Solver Derivation}
\label{sec:derivation}
\subsection{From Linear Algebra to Probabilistic Inference}
We begin our derivation by defining an undirected graphical model
(\ie, a Markov random field), $\mathcal{G}$, corresponding to the
linear system of equations. Specifically, let
$\mathcal{G}=(\mathcal{X},\mathcal{E})$, where $\mathcal{X}$ is a
set of nodes that are in one-to-one correspondence with the linear
system's variables $\vx=\{x_{1},\ldots,x_{n}\}^{T}$, and where
$\mathcal{E}$ is a set of undirected edges determined by the
non-zero entries of the (symmetric) matrix $\mA$.

Using this graph, we can translate the problem of solving the
linear system from the algebraic domain to the domain of
probabilistic inference, as stated in the following theorem.

\begin{prop}\label{prop_3}
The computation of the solution vector $\vx^{\ast}$ is identical
to the inference of the vector of marginal means
\mbox{$\mathbf{\mu}\triangleq\{\mu_{1},\ldots,\mu_{n}\}$} over the
graph $\mathcal{G}$ with the associated joint Gaussian probability
density function  $p(\vx)\sim\mathcal{N}(\mu,\mA^{-1})$.
\end{prop}
\begin{proof}[{\bf Proof}]
Another way of solving the set of linear equations
$\mA\vx-\vb=\mathbf{0}$ is to represent it by using a quadratic
form \mbox{$q(\vx)\triangleq\vx^{T}\mA\vx/2-\vb^{T}\vx$}. As the
matrix $\mA$ is symmetric, the derivative of the quadratic form
w.r.t. the vector $\vx$ is given by the vector $\partial
q/\partial\vx=\mA\vx-\vb$. Thus equating $\partial q/\partial
\vx=\mathbf{0}$ gives the global minimum $\vx^{\ast}$ of this
convex function, which is nothing but the desired solution to
$\mA\vx=\vb$.

Next, one can define the following joint Gaussian probability
density function \BE\label{eq_G}
p(\vx)\triangleq\mathcal{Z}^{-1}\exp{\big(-q(\vx)\big)}=\mathcal{Z}^{-1}\exp{(-\vx^{T}\mA\vx/2+\vb^{T}\vx)},\EE
where $\mathcal{Z}$ is a distribution normalization factor.
Denoting the vector $\mathbf{\mu}\triangleq\mA^{-1}\vb$, the
Gaussian density function can be rewritten as \BEA\label{eq_G2}
p(\vx)&=&\mathcal{Z}^{-1}\exp{(\mathbf{\mu}^{T}\mA\mathbf{\mu}/2)}\nonumber\\&\times&\exp{(-\vx^{T}\mA\vx/2+\mathbf{\mu}^{T}\mA\vx-\mathbf{\mu}^{T}\mA\mathbf{\mu}/2)}
\nonumber\\&=&\mathcal{\zeta}^{-1}\exp{\big(-(\vx-\mathbf{\mu})^{T}\mA(\vx-\mathbf{\mu})/2\big)}\nonumber\\&=&\mathcal{N}(\mathbf{\mu},\mA^{-1}),\EEA
where the new normalization factor
$\mathcal{\zeta}\triangleq\mathcal{Z}\exp{(-\mathbf{\mu}^{T}\mA\mathbf{\mu}/2)}$.
It follows that the target solution $\vx^{\ast}=\mA^{-1}\vb$ is
equal to $\mathbf{\mu}\triangleq\mA^{-1}\vb$, the mean vector of
the distribution $p(\vx)$, as defined above~(\ref{eq_G}).

Hence, in order to solve the system of linear equations we need to
infer the marginal densities, which must also be Gaussian,
\mbox{$p(x_{i})\sim\mathcal{N}(\mu_{i}=\{\mA^{-1}\vb\}_{i},P_{i}^{-1}=\{\mA^{-1}\}_{ii})$},
where $\mu_{i}$ and $P_{i}$ are the marginal mean and inverse
variance (sometimes called the precision), respectively.
\end{proof}

According to Proposition~\ref{prop_3}, solving a deterministic
vector-matrix linear equation translates to solving an inference
problem in the corresponding graph. The move to the probabilistic
domain calls for the utilization of BP as an efficient inference
engine.

\begin{remark}
Defining a jointly Gaussian probability density function,
immediately yields an implicit assumption on the positive
semi-definiteness of the precision matrix $\mA$, in addition to
the symmetry assumption. However, we would like to stress out that
this assumption emerges only for exposition purposes, so we can
use the notion of `Gaussian probability', but the derivation of
the GaBP solver itself does not use this assumption. See the
numerical example of the exact GaBP-based solution of a system
with a symmetric, but not positive semi-definite, data matrix
$\mA$ in Section~\ref{sec_nonPSD}.
\end{remark}
\subsection{Belief Propagation}
Belief propagation (BP) is equivalent to applying Pearl's local
message-passing algorithm~\cite{BibDB:BookPearl}, originally
derived for exact inference in trees, to a general graph even if
it contains cycles (loops). BP has been found to have outstanding
empirical success in many applications, \eg, in decoding Turbo
codes and low-density parity-check (LDPC) codes. The excellent
performance of BP in these applications may be attributed to the
sparsity of the graphs, which ensures that cycles in the graph are
long, and inference may be performed as if it were a tree.

The BP algorithm functions by passing real-valued messages across
edges in the graph and consists of two computational rules, namely
the `sum-product rule' and the `product rule'. In contrast to
typical applications of BP in coding theory~\cite{BibDB:AjiMcEliece}, our graphical
representation resembles to a pairwise Markov random
field\cite{BibDB:BookJordan} with a single type of propagating
messages, rather than a factor graph~\cite{BibDB:FactorGraph} with
two different types of messages, originated from either the
variable node or the factor node. Furthermore, in most graphical
model representations used in the information theory literature
the graph nodes are assigned with discrete values, while in this
contribution we deal with nodes corresponding to continuous
variables. Thus, for a graph $\mathcal{G}$ composed of potentials
$\psi_{ij}$ and $\phi_{i}$ as previously defined, the conventional
sum-product rule becomes an integral-product
rule~\cite{BibDB:Weiss01Correctness} and the message
$m_{ij}(x_j)$, sent from node $i$ to node $j$ over their shared
edge on the graph, is given by \BE\label{eq_contBP}
    m_{ij}(x_j)\propto\int_{x_i} \psi_{ij}(x_i,x_j) \phi_{i}(x_i)
\prod_{k \in \textrm{N}(i)\setminus j} m_{ki}(x_i) dx_{i}. \EE The
marginals are computed (as usual) according to the product rule
\BE\label{eq_productrule}
p(x_{i})=\alpha
\phi_{i}(x_{i})\prod_{k\in\textrm{N}(i)}m_{ki}(x_{i}), \EE where
the scalar $\alpha$  is a normalization constant. Note that the
propagating messages (and the graph potentials) do not have to
describe valid (\ie, normalized) density probability functions, as
long as the inferred marginals do.

\subsection{The Gaussian BP Algorithm}\label{sec:GaBP}
Gaussian BP is a special case of continuous BP, where the
underlying distribution is Gaussian. Now, we derive the Gaussian
BP update rules by substituting Gaussian distributions into the
continuous BP update equations.

Given the data matrix $\mA$ and the observation vector $\vb$, one
can write explicitly the Gaussian density function,
$p(\vx)$~(\ref{eq_G2}), and its corresponding graph $\mathcal{G}$.
Using the graph definition and a certain (arbitrary) pairwise
factorization of the Gaussian function~(\ref{eq_G2}), the edge
potentials (compatibility functions) and self potentials
('evidence') $\phi_{i}$ are determined to be
\BEA\psi_{ij}(x_{i},x_{j})&\triangleq&
\exp(-x_{i}A_{ij}x_{j}),\\
\phi_{i}(x_{i})&\triangleq&\exp\big(b_{i}x_{i}-A_{ii}x_{i}^{2}/2\big),\EEA
respectively. Note that by completing the square, one can observe
that \BE\label{eq_PhiGauss}
\phi_{i}(x_{i})\propto\mathcal{N}(\mu_{ii}=b_{i}/A_{ii},P_{ii}^{-1}=A_{ii}^{-1}).\EE
The graph topology is specified by the structure of the matrix
$\mA$, \ie the edges set $\{i,j\}$ includes all non-zero entries
of $\mA$ for which $i>j$.

Before describing the inference algorithm performed over the
graphical model, we make the elementary but very useful
observation that the product of Gaussian densities over a common
variable is, up to a constant factor, also a Gaussian density.

\begin{lem}\label{Lemma_Gaussian}
Let $f_{1} (x)$ and $f_{2} (x)$ be the probability density
functions of a Gaussian random variable with two possible
densities $\mathcal{N}(\mu_{1},P_{1}^{-1})$ and
$\mathcal{N}(\mu_{2},P_{2}^{-1})$, respectively. Then their
product, $f(x)=f_1(x) f_2(x)$ is, up to a constant factor, the
probability density function of a Gaussian random variable with
distribution $\mathcal{N}(\mu,P^{-1})$, where \BEA
\mu&=&P^{-1}(P_{1}\mu_{1}+P_{2}\mu_{2})\label{eq_productmean},\\
P^{-1}&=&(P_{1}+P_{2})^{-1}\label{eq_productprec}. \EEA
\end{lem}

The proof of this lemma is found in Appendix~\ref{App_Lemma}.

\begin{figure}[h!]
\begin{center}
    \includegraphics[width=0.4\textwidth]{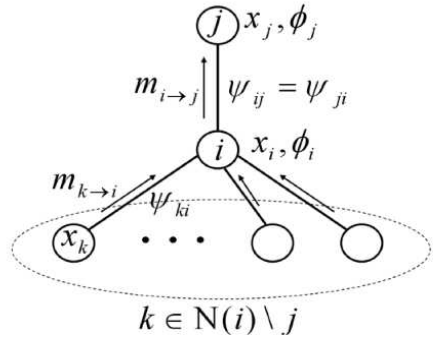}
\caption{Belief propagation message flow}
\label{fig_flow}
\end{center}
\end{figure}

Fig.~\ref{fig_flow} plots a portion of a certain graph, describing
the neighborhood of node $i$. Each node (empty circle) is
associated with a variable and self potential $\phi$, which is a
function of this variable, while edges go with the pairwise
(symmetric) potentials $\Psi$. Messages are propagating along the
edges on both directions (only the messages relevant for the
computation of $m_{ij}$ are drawn in Fig.~\ref{fig_flow}). Looking
at the right hand side of the integral-product
rule~(\ref{eq_contBP}), node $i$ needs to first calculate the
product of all incoming messages, except for the message coming
from node $j$. Recall that since $p(\vx)$ is jointly Gaussian, the
factorized self potentials
$\phi_{i}(x_i)\propto\mathcal{N}(\mu_{ii},P_{ii}^{-1})$
(\ref{eq_PhiGauss}) and similarly all messages
$m_{ki}(x_i)\propto\mathcal{N}(\mu_{ki},P_{ki}^{-1})$ are of
Gaussian form as well.

As the terms in the product of the incoming messages and the self
potential in the integral-product rule~(\ref{eq_contBP}) are all a
function of the same variable, $x_{i}$ (associated with the node
$i$), then, according to the multivariate extension of
Lemma~\ref{Lemma_Gaussian}, \BE\label{eq_product} \phi_{i}(x_i)
\prod_{k \in \textrm{N}(i) \backslash j} m_{ki}(x_i) \EE is
proportional to a certain Gaussian distribution,
$\mathcal{N}(\mu_{i\backslash j},P_{i\backslash j}^{-1})$.
Applying the multivariate version of the product precision
expression in~(\ref{eq_productprec}), the update rule for the
inverse variance is given by (over-braces denote the origin of
each of the terms) \BE\label{eq_prec} P_{i\backslash j} =
\overbrace{P_{ii}}^{\phi_{i}(x_i)} + \sum_{k\in \textrm{N}(i)
\backslash j} \overbrace{P_{ki}}^{m_{ki}(x_i)}, \EE where
$P_{ii}\triangleq A_{ii}$ is the inverse variance a-priori
associated with node $i$, via the precision of $\phi_{i}(x_{i})$,
and $P_{ki}$ are the inverse variances of the messages
$m_{ki}(x_i)$. Similarly using~(\ref{eq_productmean}) for the
multivariate case, we can calculate the mean \BE\label{eq_mean}
 \mu_{i\backslash j} = P_{i\backslash j}^{-1}\Big(\overbrace{P_{ii}\mu_{ii}}^{\phi_{i}(x_i)} +
\sum_{{k} \in \textrm{N}(i) \backslash j}
\overbrace{P_{ki}\mu_{ki}}^{m_{ki}(x_i)}\Big), \EE where
$\mu_{ii}\triangleq b_{i}/A_{ii}$ is the mean of the self
potential and $\mu_{ki}$ are the means of the incoming messages.

Next, we calculate the remaining terms of the message
$m_{ij}(x_j)$, including the integration over $x_{i}$. After some
algebraic manipulation (deferred to Appendix~\ref{App_Integral}),
using the Gaussian integral \BE\label{eq_GaussianIntegral}
\int_{-\infty}^{\infty}\exp{(-ax^{2}+bx)}dx=\sqrt{\pi/a}\exp{(b^{2}/4a)},
\EE we find that the messages $m_{ij}(x_j)$ are proportional to
normal distribution with precision and mean
\BE\label{eq_prec_message} P_{ij} = -A_{ij}^2P_{i \backslash
j}^{-1}, \EE \BE \mu_{ij}\label{eq_mean_message} =
-P_{ij}^{-1}A_{ij}\mu_{i\backslash j}. \EE These two scalars
represent the messages propagated in the Gaussian BP-based
algorithm.
Finally, computing the product rule~(\ref{eq_productrule}) is
similar to the calculation of the previous
product~(\ref{eq_product}) and the resulting mean~(\ref{eq_mean})
and precision~(\ref{eq_prec}), but including all incoming
messages. The marginals are inferred by normalizing the result of
this product. Thus, the marginals are found to be Gaussian
probability density functions $\mathcal{N}(\mu_{i},P_{i}^{-1})$
with precision and mean \BEA P_{i}
= \overbrace{P_{ii}}^{\phi_{i}(x_i)} + \sum_{k\in \textrm{N}(i)} \overbrace{P_{ki}}^{m_{ki}(x_i)},\\
\mu_{i} = P_{i\backslash
j}^{-1}\Big(\overbrace{P_{ii}\mu_{ii}}^{\phi_{i}(x_i)} + \sum_{{k}
\in \textrm{N}(i)}
\overbrace{P_{ki}\mu_{ki}}^{m_{ki}(x_i)}\Big)\label{eq_marginal_mean},
\EEA respectively. The derivation of the GaBP-based solver
algorithm is concluded by simply substituting the explicit derived expressions
of $P_{i\backslash j}$~(\ref{eq_prec}) into
$P_{ij}$~(\ref{eq_prec_message}), $\mu_{i\backslash
j}$~(\ref{eq_mean}) and $P_{ij}$~(\ref{eq_prec_message}) into
$\mu_{ij}$~(\ref{eq_mean_message}) and $P_{i\backslash
j}$~(\ref{eq_prec}) into $\mu_{i}$~(\ref{eq_marginal_mean}).

The message passing in the GaBP solver can be performed subject to
any scheduling. We refer to two conventional messages updating
rules: parallel (flooding or synchronous) and serial (sequential,
asynchronous) scheduling. In the parallel scheme, messages are
stored in two data structures: messages from the previous
iteration round, and messages from the current round. Thus,
incoming messages do not affect the result of the computation in
the current round, since it is done on messages that were received
in the previous iteration round. Unlike this, in the serial
scheme, there is only one data structure, so incoming messages in
this round, change the result of the computation. In a sense it is
exactly like the difference between the Jacobi and Guess-Seidel
algorithms, to be discussed in the following. Some in-depth
discussions about parallel vs. serial scheduling in the BP context
can be found in the work of Elidan \etal~\cite{BibDB:ElidanEtAl}.

\begin{table}[h!]
\begin{alg}\label{alg_GaBP}\end{alg}\centerline{
\begin{tabular}{|lll|}
  \hline&&\\
  \texttt{1.} & \emph{\texttt{Initialize:}} & $\checkmark$\quad\texttt{Set the neighborhood} $\textrm{N}(i)$ \texttt{to include}\\&&\quad\quad$\forall k\neq i \exists A_{ki}\neq0$.\\&& $\checkmark$\quad\texttt{Set the scalar fixes}\\&&$\quad\quad P_{ii}=A_{ii}$ \texttt{and} $\mu_{ii}=b_{i}/A_{ii}$, $\forall i$.\\
  && $\checkmark$\quad\texttt{Set the initial $\textrm{N}(i)\ni k\rightarrow i$ scalar messages}\\&&\quad\quad $P_{ki}=0$ \texttt{and} $\mu_{ki}=0$.\\&& $\checkmark$\quad \texttt{Set a convergence threshold} $\epsilon$.\\
  {\texttt{2.}} & {\emph{\texttt{Iterate:}} }&  $\checkmark$\quad\texttt{Propagate the $\textrm{N}(i)\ni k\rightarrow i$ messages}\\&&$\quad\quad P_{ki}$ \texttt{and} $\mu_{ki}$, $\forall i$ \texttt{(under certain scheduling)}.\\&&
  $\checkmark$\quad\texttt{Compute the} $\textrm{N}(j)\ni i\rightarrow j$ \texttt{scalar messages} \\&& $\quad\quad P_{ij} = -A_{ij}^{2}/\big(P_{ii}+\sum_{{k}\in\textrm{N}(i) \backslash j}
P_{ki}\big)$,\\&& $\quad\quad\mu_{ij} =
\big(P_{ii}\mu_{ii}+\sum_{k \in \textrm{N}(i) \backslash j}
P_{ki}\mu_{ki}\big)/A_{ij}$.\\
  {\texttt{3.}} & {\emph{\texttt{Check:}}} & $\checkmark$\quad\texttt{If the messages} $P_{ij}$ \texttt{and} $\mu_{ij}$ \texttt{did not}\\&&\quad\quad\texttt{converge (w.r.t. $\epsilon$),} \texttt{return to
    Step 2.}\\&&$\checkmark$\quad\texttt{Else, continue to Step 4.}\\
  {\texttt{4.}} & {\emph{\texttt{Infer:}}} & $\checkmark$\quad\texttt{Compute the marginal means}\\&&{\quad\quad$\mu_{i}=\big(P_{ii}\mu_{ii}+\sum_{k \in
\textrm{N}(i)}P_{ki}\mu_{ki}\big)/\big(P_{ii}+\sum_{{k}\in\textrm{N}(i)}
P_{ki}\big)$, $\forall i$.}\\&&$(\checkmark$\quad\texttt{Optionally compute the marginal precisions}\\&&\quad\quad$P_{i}=P_{ii}+\sum_{k\in\textrm{N}(i)}P_{ki}\quad)$\\
  {\texttt{5.}} & {\emph{\texttt{Solve:}}} & $\checkmark$\quad\texttt{Find the solution}\\&& {$\quad\quad x_{i}^{\ast}=\mu_{i}$, $\forall i$.}\\&&\\\hline
\end{tabular}
}\newpage
\end{table}
\subsection{Max-Product Rule}
\label{MaxProductRule}
A well-known alternative version to the sum-product BP is the
max-product (a.k.a. min-sum) algorithm~\cite{Max-product}. In this variant of
BP, maximization operation is performed rather than
marginalization, \ie, variables are eliminated by taking maxima
instead of sums. For Trellis trees (\eg, graphically representing
convolutional codes or ISI channels), the conventional sum-product
BP algorithm boils down to performing the BCJR
algorithm\cite{BibDB:BCJR}, resulting in the most probable symbol,
while its max-product counterpart is equivalent to the Viterbi
algorithm~\cite{Viterbi}, thus inferring the most probable sequence of
symbols~\cite{BibDB:FactorGraph}.

In order to derive the max-product version of the proposed GaBP
solver, the integral(sum)-product rule~(\ref{eq_contBP}) is
replaced by a new rule\BE\label{eq_contBP2}
    m_{ij}(x_j)\propto\argmax_{x_i} \psi_{ij}(x_i,x_j) \phi_{i}(x_i)
\prod_{k \in \textrm{N}(i)\setminus j} m_{ki}(x_i) dx_{i}. \EE
Computing $m_{ij}(x_j)$ according to this max-product rule, one
gets (the exact derivation is deferred to Appendix~\ref{App_Max})
\BE
m_{ij}(x_j)\propto\mathcal{N}(\mu_{ij}=-P_{ij}^{-1}A_{ij}\mu_{i\backslash
j},P_{ij}^{-1} = -A_{ij}^{-2}P_{i \backslash j}), \EE which is
identical to the messages derived for the sum-product
case~(\ref{eq_prec_message})-(\ref{eq_mean_message}). Thus
interestingly, as opposed to ordinary (discrete) BP, the following
property of the GaBP solver emerges.
\begin{corol}
The max-product (Eq.~\ref{eq_contBP2}) and sum-product (Eq. ~\ref{eq_contBP})
versions of the GaBP solver are identical.
\end{corol}

\section{Convergence and Exactness}
\label{sec:converge-exact} In ordinary BP, convergence does not
entail exactness of the inferred probabilities, unless the graph
has no cycles.   Luckily, this is not the case for the GaBP
solver. Its underlying Gaussian nature yields a direct connection
between convergence and exact inference. Moreover, in contrast to
BP the convergence of GaBP is not limited for tree or sparse
graphs and can occur even for dense (fully-connected) graphs,
adhering to certain rules discussed in the following.

We can use results from the literature on probabilistic inference
in graphical
models~\cite{BibDB:Weiss01Correctness,BibDB:jmw_walksum_nips,BibDB:mjw_walksum_jmlr06}
to determine the convergence and exactness properties of the
GaBP-based solver. The following two theorems establish sufficient
conditions under which GaBP is guaranteed to converge to the exact
marginal means.

\begin{thm}{\cite[Claim 4]{BibDB:Weiss01Correctness}}
If the matrix $\mA$ is strictly diagonally dominant, then GaBP
converges and the marginal means converge to the true means.
\end{thm}

This sufficient condition was recently relaxed to include a wider
group of matrices.

\begin{thm}{\cite[Proposition 2]{BibDB:jmw_walksum_nips}}
\label{spectral_radius_thm}
If the spectral radius of the matrix $\mA$ satisfies \BE
\rho(|\mI_{n}-\mA|)<1, \EE then GaBP converges and the marginal
means converge to the true means. (The assumption here is that the matrix $\mA$ is first
normalized by multiplying with $\mD^{-1}$, where $\mD = diag(\mA)$.)
\end{thm}

A third and weaker sufficient convergence condition (relative to Theorem \ref{spectral_radius_thm})
which characterizes the convergence of the variances is given in
\cite[Theorem 2]{MinSum}: For each row in the matrix $\mA$, if $A_{ii}^2 > \Sigma_{j \ne i} A_{ij}^2$ then
the variances converge. Regarding the means, additional condition related to Theorem \ref{spectral_radius_thm} is given. \\

There are many examples of linear systems that violate these
conditions, for which GaBP converges to the exact means. In
particular, if the graph corresponding to the system is acyclic
(\ie, a tree), GaBP yields the exact marginal means (and
variances~\cite{BibDB:Weiss01Correctness}), regardless of the
value of the spectral radius of the matrix~\cite{BibDB:Weiss01Correctness}. Another example, where the graph
is fully-connected, is discussed in the following section.
However, in contrast to conventional iterative methods derived
from linear algebra, understanding the conditions for exact
convergence and quantifying the convergence rate of the GaBP
solver remain intriguing open problems.

\subsection{Convergence Acceleration}
Further speed-up of GaBP can
be achieved by adapting known acceleration techniques from linear
algebra, such Aitken's method and Steffensen's
iterations~\cite{BibDB:BookHenrici}. Consider a sequence
$\{x_{n}\}$ (\eg, obtained by using GaBP iterations) linearly
converging to the limit $\hat{x}$, and $x_n \ne \hat{x}$ for $n
\ge 0$. According to Aitken's method, if there exists a real
number $a$ such that $|a|<1 $ and \mbox{$\lim_{n \rightarrow
\infty}(x_n-\hat{x})/(x_{n-1} - \hat{x}) = a$}, then the sequence
$\{ y_n\}$ defined by
\[ y_n = x_n - \frac{(x_{n+1} -x_n)^2}{x_{n+2} - 2x_{n+1} + x_n} \]
converges to $\hat{x}$ faster than $\{ x_n \}$ in the sense that
\mbox{$\lim_{n \rightarrow \infty} |(\hat{x} - y_n)/(\hat{x} -
x_n)| = 0$}. Aitken's method can be viewed as a generalization of
over-relaxation, since one uses values from three, rather than
two, consecutive iteration rounds. This method can be easily
implemented in GaBP as every node computes values based only on
its own history.

Steffensen's iterations incorporate Aitken's method. Starting with
$x_{n}$, two iterations are run to get $x_{n+1}$ and $x_{n+2}$.
Next, Aitken's method is used to compute $y_{n}$, this value
replaces the original $x_{n}$, and GaBP is executed again to get a
new value of $x_{n+1}$. This process is repeated iteratively until
convergence. \comment{Table~\ref{tab_2} demonstrates the speed-up
of GaBP obtained by using these acceleration methods, in
comparison with that achieved by the similarly modified Jacobi
algorithm.\footnote{Application of Aitken and Steffensen's methods
for speeding-up the convergence of standard (non-BP) iterative
solution algorithms in the context of MUD was introduced by Leibig
\etal~\cite{LDF05}.}} We remark that, although the convergence
rate is improved with these enhanced algorithms, the region of
convergence of the accelerated GaBP solver remains unchanged.

\comment{
\subsection{Approximation for Asymmetric Data Matrices}\label{sec_asymmetric}

\footnote{For the case of  the approximate solution of asymmetric
linear systems is discussed in Section~\ref{sec_asymmetric}}.

if we want to keep sparsity. }

\section{Computational Complexity and Message-Passing Efficiency}
\label{sec:complexity}
\comment{For a dense data matrix $\mA$, The GaBP solver
algorithm~\ref{alg_GaBP} can be easily implemented in a distributed
fashion. Each node $i$ on the corresponding graph receives as an
input the $i$'th row (or column) of the symmetric data matrix
$\mA$ and the scalar observation $b_{i}$. In each iteration, for
every non-zero entry $A_{ij}$, a message containing two real
numbers, $\mu_{ij}$ and $P_{ij}$, is sent from node $i$ to node
$j$ along their shared edge.

For a dense matrix $\mA$ each node out of the $n$ nodes sends a
unique message to every other node on the fully-connected graph.
This recipe results in a total of $\mathcal{o}(n^2)$ messages per
iteration round.

The algorithm can be easily implemented in a distributed fashion.
Each node $i$ on the corresponding graph receives as an input the
$i$'th row (or column) of the data matrix $\mA$ and the scalar
observation $b_{i}$. In each iteration, for every non-zero entry
$A_{ij}$, a message containing two real numbers, $\mu_{ij}$ and
$P_{ij}$, is sent from node $i$ to node $j$ along their shared
edge.}

For a dense matrix $\mA$ each
node out of the $n$ nodes sends a unique message to every other
node on the fully-connected graph. This recipe results in a total
of $n^{2}$ messages per iteration round.

The computational complexity of the GaBP solver as described in
Algorithm~\ref{alg_GaBP} for a dense linear system, in terms of
operations (multiplications and additions) per iteration round. is
shown in Table~\ref{tab_complexity}. In this case, the total
number of required operations per iteration is
$\mathcal{O}(n^{3})$. This number is obtained by evaluating the
number of operations required to generate a message multiplied by
the number of messages. Based on the summation expressions for the
propagating messages $P_{ij}$ and $\mu_{ij}$, it is easily seen
that it takes $\mathcal{O}(n)$ operations to compute such a
message. In the dense case, the graph is fully-connected resulting
in $\mathcal{O}(n^{2})$ propagating messages.

In order to estimate the total number of operations required for
the GaBP algorithm to solve the linear system, we have to evaluate
the number of iterations required for convergence. It is
known~\cite{BibDB:BookBertsekasTsitsiklis} that the number of iterations required for an
iterative solution method is $\mathcal{O}(f(\kappa))$, where
$f(\kappa)$ is a function of the condition number of the data
matrix $\mA$. Hence the total complexity of the GaBP solver can be
expressed by $\mathcal{O}(n^3)\times\mathcal{O}(f(\kappa))$. The
analytical evaluation of the convergence rate function $f(\kappa)$
is a challenging open problem. However, it can be upper bounded by
$f(\kappa)<\kappa$. furthermore, based on our experimental study,
described in Section~\ref{sec_results}, we can conclude that
$f(\kappa)\leq\sqrt{\kappa}$. Thus, the total complexity of the
GaBP solve in this case is
$\mathcal{O}(n^{3})\times\mathcal{O}(\sqrt{\kappa})$. For
well-conditioned (as opposed to ill-conditioned) data matrices the
condition number is $\mathcal{O}(1)$. Thus, for well-conditioned
linear systems the total complexity is $\mathcal{O}(n^3)$, \ie,
the complexity is cubic, the same order as for direct solution
methods, like Gaussian elimination.

At first sight, this result may be considered disappointing, with
no complexity gain w.r.t. direct matrix inversion. Luckily, the
GaBP implementation as described in Algorithm~\ref{alg_GaBP} is a
naive one, thus termed naive GaBP. In this implementation we did
not take into account the correlation between the different
messages transmitted from a certain node $i$. These messages,
computed by summation, distinct from one another in only two
summation terms.

\begin{table}
\centerline{
\begin{tabular}{|l|c|c|c|}
  \hline
  \textbf{Algorithm}
  & \textbf{Operations per msg} & \textbf{msgs} & \textbf{Total operations per iteration}\\\hline\hline & & &\\
  Naive GaBP (Algorithm~\ref{alg_GaBP}) & $\mathcal{O}(n)$ & $\mathcal{O}(n^{2})$ & $\mathcal{O}(n^{3})$ \\\hline & & & \\
  Broadcast GaBP (Algorithm~\ref{alg_GaBP_Broadcast}) & $\mathcal{O}(n)$ & $\mathcal{O}(n)$ & $\mathcal{O}(n^{2})$\\
  \hline
\end{tabular}
}\vspace{0.5cm} \caption{Computational complexity of the GaBP
solver for dense $n\times n$ matrix $\mA$.}\label{tab_complexity}
\end{table}

\comment{The number of messages passed on the graph can be reduced
significantly down to $n$ messages per round by using a similar
construction to Bickson \etal~\cite{BroadcastBP}. This
modification is described in the following corollary:
\begin{corol}[GaBP solver with reduced message passing]\label{corol_5}
Instead of sending a message composed of the pair of $\mu_{ij}$
and $P_{ij}$, a node broadcasts aggregated sums, and consequently
each node can retrieve locally the $P_{i\backslash
j}$~(\ref{eq_prec}) and $\mu_{i\backslash j}$~(\ref{eq_mean}) from
the sums by means of a subtraction:
\end{corol}}

\begin{table}[h!]
\begin{alg}\label{alg_GaBP_Broadcast}\end{alg}\centerline{
\begin{tabular}{|lll|}
  \hline&&\\
  \texttt{1.} & \emph{\texttt{Initialize:}} & $\checkmark$\quad\texttt{Set the neighborhood} $\textrm{N}(i)$ \texttt{to include}\\&&\quad\quad$\forall k\neq i \exists A_{ki}\neq0$.\\&& $\checkmark$\quad\texttt{Set the scalar fixes}\\&&$\quad\quad P_{ii}=A_{ii}$ \texttt{and} $\mu_{ii}=b_{i}/A_{ii}$, $\forall i$.\\
  && $\checkmark$\quad\texttt{Set the initial $i\rightarrow\textrm{N}(i)$ broadcast messages}\\&&\quad\quad $\tilde{P_{i}}=0$ \texttt{and} $\tilde{\mu}_{i}=0$.\\&&$\checkmark$\quad\texttt{Set the initial $\textrm{N}(i)\ni k\rightarrow i$ internal scalars}\\&&\quad\quad $P_{ki}=0$ \texttt{and} $\mu_{ki}=0$.\\&& $\checkmark$\quad \texttt{Set a convergence threshold} $\epsilon$.\\
  {\texttt{2.}} & {\emph{\texttt{Iterate:}} }&  $\checkmark$\quad\texttt{Broadcast the aggregated sum messages}\\&&$\quad\quad \tilde{P}_{i}=P_{ii}+\sum_{{k}\in\textrm{N}(i)}
P_{ki}$,\\&&
$\quad\quad\tilde{\mu}_{i}=\tilde{P_{i}}^{-1}(P_{ii}\mu_{ii}+\sum_{k
\in \textrm{N}(i)} P_{ki}\mu_{ki})$, $\forall
i$\\&&\quad\quad\texttt{(under certain scheduling)}.\\&&
  $\checkmark$\quad\texttt{Compute the} $\textrm{N}(j)\ni i\rightarrow j$ \texttt{internal scalars} \\&& $\quad\quad P_{ij} = -A_{ij}^{2}/(\tilde{P}_{i}-P_{ji})$,\\
  &&$\quad\quad\mu_{ij}=(\tilde{P_{i}}\tilde{\mu_{i}}-P_{ji}\mu_{ji})/A_{ij}$.\\
  {\texttt{3.}} & {\emph{\texttt{Check:}}} & $\checkmark$\quad\texttt{If the internal scalars} $P_{ij}$ \texttt{and} $\mu_{ij}$ \texttt{did not}\\&&\quad\quad\texttt{converge (w.r.t. $\epsilon$),} \texttt{return to
    Step 2.}\\&&$\checkmark$\quad\texttt{Else, continue to Step 4.}\\
  {\texttt{4.}} & {\emph{\texttt{Infer:}}} & $\checkmark$\quad\texttt{Compute the marginal means}\\&&{\quad\quad$\mu_{i}=\big(P_{ii}\mu_{ii}+\sum_{k \in
\textrm{N}(i)}P_{ki}\mu_{ki}\big)/\big(P_{ii}+\sum_{{k}\in\textrm{N}(i)}
P_{ki}\big)=\tilde{\mu}_{i}$, $\forall i$.}\\&&$(\checkmark$\quad\texttt{Optionally compute the marginal precisions}\\&&\quad\quad$P_{i}=P_{ii}+\sum_{k\in\textrm{N}(i)}P_{ki}=\tilde{P}_{i}\quad)$\\
  {\texttt{5.}} & {\emph{\texttt{Solve:}}} & $\checkmark$\quad\texttt{Find the solution}\\&& {$\quad\quad x_{i}^{\ast}=\mu_{i}$, $\forall i$.}\\&&\\\hline
\end{tabular}}
\end{table}

Instead of sending a message composed of the pair of $\mu_{ij}$
and $P_{ij}$, a node can broadcast the aggregated sums \BEA
\tilde{P}_{i}&=&P_{ii}+\sum_{{k}\in\textrm{N}(i)}
P_{ki},\\\tilde{\mu}_{i}&=&\tilde{P}_{i}^{-1}(P_{ii}\mu_{ii}+\sum_{k
\in \textrm{N}(i)} P_{ki}\mu_{ki}). \EEA Consequently, each node
can retrieve locally the $P_{i\backslash j}$~(\ref{eq_prec}) and
$\mu_{i\backslash j}$~(\ref{eq_mean}) from the sums by means of a
subtraction \BEA P_{i\backslash
j}&=&\tilde{P}_{i}-P_{ji},\\\mu_{i\backslash
j}&=&\tilde{\mu}_{i}-P_{i \backslash j}^{-1}P_{ji}\mu_{ji}.\EEA
The rest of the algorithm remains the same. On dense graphs, the broadcast version sends
$O(n)$ messages per round, instead of $O(n^2)$ messages in the GaBP algorithm.

\section{The GaBP-Based Solver and Classical Solution Methods}
\label{sec:classical}
\subsection{Gaussian Elimination}

\begin{prop}\label{prop_GaBPGE}
The GaBP-based solver (Algorithm~\ref{alg_GaBP}) for a system of
linear equations represented by a tree graph is identical to the
renowned Gaussian elimination algorithm (a.k.a. LU
factorization,~\cite{BibDB:BookBertsekasTsitsiklis}).
\end{prop}
\begin{proof}[{\bf Proof}]
Consider a set of $n$ linear equations with $n$ unknown variables,
a unique solution and a tree graph representation. We aim at
computing the unknown variable associated with the root node.
Without loss of generality as the tree can be drawn with any of
the other nodes being its root. Let us enumerate the nodes in an
ascending order from the root to the leaves (see, e.g.,
Fig. \ref{fig_tree}).

\begin{center}
\begin{figure}[h!]
\begin{center}
  \includegraphics[width=100pt]{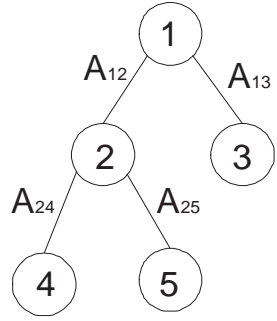}\\
  \caption{Example topology of a tree with 5 nodes}\label{fig_tree}
\end{center}
\end{figure}
\end{center}
As in a tree each child node (\ie, all nodes but the root) has
only one parent node and based on the top-down ordering, it can be
easily observed that the tree graph's corresponding data matrix
$\mA$ must have one and only one non-zero entry in the upper
triangular portion of its columns. Moreover, for a leaf node this
upper triangular entry is the only non-zero off-diagonal entry in
the whole column. See, for example, the data matrix associated
with the tree graph depicted in Fig~\ref{fig_tree} \BE \left(
  \begin{array}{ccccc}
    A_{11} & \mathbf{A_{12}} & \underline{\mathbf{A_{13}}} & 0 & 0 \\
    A_{12} & A_{22} & 0 & \underline{\mathbf{A_{24}}} & \underline{\mathbf{A_{25}}} \\
    A_{13} & 0 & A_{33} & 0 & 0 \\
    0 & A_{24} & 0 & A_{44} & 0 \\
    0 & A_{25} & 0 & 0 & A_{55} \\
  \end{array}
\right), \EE where the non-zero upper triangular entries are in
bold and among these the entries corresponding to leaves are
underlined.

Now, according to GE we would like to lower triangulate the matrix
$\mA$. This is done by eliminating these entries from the leaves
to the root. Let $l$ be a leaf node, $i$ be its parent and $j$ be
its parent ($l$'th node grandparent). Then, the $l$'th row is
multiplied by $-A_{li}/A_{ll}$ and added to the $i$'th row. in
this way the $A_{li}$ entry is being eliminated. However, this
elimination, transforms the $i$'th diagonal entry to be
$A_{ii}\rightarrow A_{ii}-A_{li}^{2}/A_{ll}$, or for multiple
leaves connected to the same parent $A_{ii}\rightarrow
A_{ii}-\sum_{l\in\textrm{N}(i)\j}A_{li}^{2}/A_{ll}$. In our
example, \BE \left(
  \begin{array}{ccccc}
    A_{11} & A_{12} & 0 & 0 & 0 \\
    A_{12} & A_{22}-A_{13}^2/A_{33}-A_{24}^2/A_{44}-A_{25}^2/A_{55} & 0 & 0 & 0 \\
    A_{13} & 0 & A_{33} & 0 & 0 \\
    0 & A_{24} & 0 & A_{44} & 0 \\
    0 & A_{25} & 0 & 0 & A_{55} \\
  \end{array}
\right). \EE

Thus, in a similar manner, eliminating the parent $i$ yields the
multiplication of the $j$'th diagonal term by
$-A_{ij}^{2}/(A_{ii}-\sum_{l\in\textrm{N}(i)\j}A_{li}^{2}/A_{ll})$.
Recalling that $P_{ii}=A_{ii}$, we see that the last expression is
identical to the update rule of $P_{ij}$ in GaBP. Again, in our
example \BE \left(
  \begin{array}{ccccc}
    B & 0 & 0 & 0 & 0 \\
    0 & C & 0 & 0 & 0 \\
    A_{13} & 0 & A_{33} & 0 & 0 \\
    0 & A_{24} & 0 & A_{44} & 0 \\
    0 & A_{25} & 0 & 0 & A_{55} \\
  \end{array}
\right), \EE where $B =
A_{11}-A_{12}^{2}/(A_{22}-A_{13}^2/A_{33}-A_{24}^2/A_{44}-A_{25}^2/A_{55})$,
$C=A_{22}-A_{13}^2/A_{33}-A_{24}^2/A_{44}-A_{25}^2/A_{55}$. Now
the matrix is fully lower triangulated. To put differently in
terms of GaBP, the $P_{ij}$ messages are subtracted from the
diagonal $P_{ii}$ terms to triangulate the data matrix of the
tree. Performing the same row operations on the right hand side
column vector $\vb$, it can be easily seen that we equivalently
get that the outcome of the row operations is identical to the
GaBP solver's $\mu_{ij}$ update rule. These updadtes/row
operations can be repeated, in the general case, until the matrix
is lower triangulated.

Now, in order to compute the value of the unknown variable
associated with the root node, all we have to do is divide the
first diagonal term by the transformed value of $b_{1}$, which is
identical to the infer stage in the GaBP solver (note that by
definition all the nodes connected to the root are its children,
as it does not have parent node). In the example \BE
x_{1}^{\ast}=\frac{A_{11}-A_{12}^{2}/(A_{22}-A_{13}^2/A_{33}-A_{24}^2/A_{44}-A_{25}^2/A_{55})}{b_{11}-A_{12}/(b_{22}-A_{13}/A_{33}-A_{24}/A_{44}-A_{25}/A_{55})}
\EE
\noindent Note that the rows corresponding to leaves remain unchanged.

To conclude, in the tree graph case, the `iterative' stage (stage
2 on algorithm~\ref{alg_GaBP}) of the GaBP solver actually
performs lower triangulation of the matrix, while the `infer'
stage (stage 4) reducers to what is known as forward substitution.
Evidently, using an opposite ordering, one can get the
complementary upper triangulation and back substitution,
respectively.

\end{proof}

It is important to note, that based on this proposition, the GaBP
solver can be viewed as GE ran over an unwrapped version (\ie, a
computation tree) of a general loopy graph.

\subsection{Iterative Methods}
Iterative methods that can be expressed in the simple form \BE\label{eq_iter}
\vx^{(t)}=\mB\vx^{(t-1)}+\vc, \EE where neither the iteration matrix $\mB$
nor the vector $\vc$ depend upon the iteration number $t$, are
called stationary iterative methods. In the following, we discuss
three main stationary iterative methods: the Jacobi method, the
Gauss-Seidel (GS) method and the successive overrelaxation (SOR)
method. The GaBP-based solver, in the general case, can not be
written in this form, thus can not be categorized as a stationary
iterative method.

\begin{prop}
\cite{BibDB:BookBertsekasTsitsiklis}
Assuming $\mI-\mB$ is invertible, then the iteration~\ref{eq_iter}
converges (for any initial guess, $\vx^{(0)}$).
\end{prop}

\subsection{Jacobi Method}
The Jacobi method (Gauss, 1823, and Jacobi
1845,\cite{BibDB:BookAxelsson}), a.k.a. the simultaneous iteration
method, is the oldest iterative method for solving a square linear system of
equations $\boldsymbol{A}\vx = \vb$. The method assumes that $\forall_i\mbox{  } A_{ii} \ne 0$.  It's complexity is $\mathcal{O}(n^{2})$ per iteration. A sufficient convergence condition for the Jacobi method is that for any starting vector $\boldsymbol{x}_0$ as long as $
\rho(\boldsymbol{D}^{-1}(\boldsymbol{L}+\boldsymbol{U})) < 1.$
Where $\boldsymbol{D} = diag\{\boldsymbol{A}\}$, $\boldsymbol{L},\boldsymbol{U}$ are upper and lower triangular matrices of $\boldsymbol{A}$. A second sufficient convergence condition is that
$\boldsymbol{A}$ is diagonally dominant.

\comment{If  is a strictly diagonally dominant or an irreducibly
diagonally dominant matrix, then the associated Jacobi and
Gauss-Seidel iterations converge for any }

\comment{The "true" sufficient condition for Jacobi iteration to
converge is that the "spectral radius" of
[Graphics:Images/GaussSeidelMod_gr_221.gif]  is less than 1, where
[Graphics:Images/GaussSeidelMod_gr_222.gif] is the diagonal of
[Graphics:Images/GaussSeidelMod_gr_223.gif].}


\begin{prop}\label{prop_GaBPJ}
The GaBP-based solver (Algorithm~\ref{alg_GaBP})
\begin{enumerate}
  \item with inverse variance messages arbitrarily set to zero, \ie, $P_{ij}=0, i\in\textrm{N}(j),\forall{j}$;
  \item incorporating the message received from node $j$ when computing the message to be sent from node $i$ to node $j$, \ie replacing $k\in\textrm{N}(i)\backslash j$ with $k\in\textrm{N}(i)$,
\end{enumerate}
is identical to the Jacobi iterative method.
\end{prop}
\begin{proof}[{\bf Proof}]
Arbitrarily setting the precisions to zero, we get in
correspondence to the above derivation, \BEA
P_{i\backslash j}&=&P_{ii}=A_{ii},\\
P_{ij}\mu_{ij}&=&-A_{ij}\mu_{i\backslash j},\\
\label{eq_marginal_J}\mu_{i}&=&A_{ii}^{-1}(b_{i}-\sum_{k\in\textrm{N}(i)}A_{ki}\mu_{k\backslash
i}). \EEA Note that the inverse relation between $P_{ij}$ and
$P_{i\backslash j}$~(\ref{eq_prec_message}) is no longer valid in
this case.

Now, we rewrite the mean $\mu_{i\backslash j}$~(\ref{eq_mean})
without excluding the information from node $j$, \BE
\mu_{i\backslash
j}=A_{ii}^{-1}(b_{i}-\sum_{k\in\textrm{N}(i)}A_{ki}\mu_{k\backslash
i}). \EE Note that $\mu_{i\backslash j}=\mu_{i}$, hence
the inferred marginal mean $\mu_{i}$~(\ref{eq_marginal_J}) can be
rewritten as \BE \mu_{i}=A_{ii}^{-1}(b_{i}-\sum_{k\neq
i}A_{ki}\mu_{k}), \EE where the expression for all neighbors of
node $i$ is replaced by the redundant, yet identical, expression
$k\neq i$. This fixed-point iteration is identical to the renowned
Jacobi method, concluding the proof.
\end{proof}

Proposition~\ref{prop_GaBPJ} can be viewed also as a probabilistic
proof of Jacobi. The fact that Jacobi iterations can be obtained
as a special case of the GaBP solver further indicates the
richness of the proposed algorithm. Note, that the GaBP algorithm converges to the
exact solution also for nonsymmetric matrices in this form.

\subsection{Gauss-Seidel}

The Gauss-Seidel method converges for any
starting vector $ \boldsymbol{x}_0$ if $
\rho((\boldsymbol{L}+\boldsymbol{D})^{-1}\boldsymbol{U}) < 1$.

This condition holds, for example, for diagonally dominant
matrices as well as for positive definite ones. It is necessary, however, that the diagonal terms in the matrix
are greater (in magnitude) than the other terms.

The successive overrelaxation (SOR) method aims to further refine
the Gauss-Seidel method, by adding a damping parameter $0 < \alpha
< 1$: \BE x_i^t = \alpha x_i^{t-1} + (1 - \alpha) GS_i, \EE where
$GS_i$ is the Gauss-Seidel update computed by node $i$.
Damping has previously shown to be a heuristic for accelerating
belief propagation as well \cite{Damping}.

\section{Distributed Iterative Computation of Moore-Penrose Pseudoinverse}\label{sec_pseudo}
\label{sec_new_const} In this section, we efficiently extend the
applicability of the proposed GaBP-based solver for systems with
symmetric matrices to systems with any square
(\ie, also nonsymmetric) or rectangular matrix. We first construct
a new symmetric data matrix $\tilde{\mR}$ based on an arbitrary
(non-rectangular) matrix $\mS\in\mathbb{R}^{k\times n}$ \BE
\label{newR} \tilde{\mR}\triangleq\left(
  \begin{array}{cc}
    \mI_{k} & \mS^T \\
    \mS & -\Psi \\
  \end{array}
\right)\in\mathbb{R}^{(k+n)\times(k+n)}. \EE Additionally, we
define a new vector of variables
$\tilde{\vx}\triangleq\{\hat{\vx}^{T},\vz^{T}\}^{T}\in\mathbb{R}^{(k+n)\times1}$,
where $\hat{\vx}\in\mathbb{R}^{k\times1}$ is the (to be shown)
solution vector and $\vz\in\mathbb{R}^{n\times1}$ is an auxiliary
hidden vector, and a new observation vector
$\tilde{\vy}\triangleq\{\mathbf{0}^{T},\vy^{T}\}^{T}\in\mathbb{R}^{(k+n)\times1}$.

Now, we would like to show that solving the symmetric linear
system $\tilde{\mR}\tilde{\vx}=\tilde{\vy}$, taking the first $k$
entries of the corresponding solution vector $\tilde{\vx}$
is equivalent to solving the original (not necessarily symmetric)
system $\mR\vx=\vy$. Note that in the new construction the matrix
$\tilde{\mR}$ is sparse again, and has only $2nk$ off-diagonal nonzero elements. When running the GaBP algorithm we have only $2nk$
messages, instead of $n^2$ in the previous construction.

Writing explicitly the symmetric linear system's equations, we get
\[
    \hat{\vx}+\mS^T\vz=\mathbf{0}, \]
\[    \mS\hat{\vx}-\Psi \vz=\vy.
    \]
Thus, \[ \hat{\vx}=\Psi\mS^{T}(\vy-\mS\hat{\vx}), \]
and extracting $\hat{\vx}$ we have \[
\hat{\vx}=(\mS^{T}\mS+\Psi_{n})^{-1}\mS^{T}\vy. \] Note, that
when the noise level is zero, $\Psi=0_{m \times m}$, we get the
Moore-Penrose pseudoinverse solution\[
\hat{\vx}=(\mS^{T}\mS)^{-1}\mS^{T}\vy=\mS^{\dag}\vy. \]

\section{Numerical Examples and Applications}\label{sec_results}
Our experimental study includes four numerical examples and two
possible applications. In all examples, but the Poisson's
equation~\ref{eq_poisson}, $\vb$ is assumed to be an $m$-length all-ones
observation vector. For fairness in comparison, the initial guess
in all experiments, for the various solution methods under
investigation, is taken to be the same and is arbitrarily set to
be equal to the value of the vector $\vb$. The stopping criterion
in all experiments determines that for all propagating messages
(in the context the GaBP solver) or all $n$ tentative solutions
(in the context of the compared iterative methods) the absolute
value of the difference should be less than $\epsilon\leq10^{-6}$.
As for terminology, in the following performing GaBP with parallel
(flooding or synchronous) message scheduling is termed `parallel
GaBP', while GaBP with serial (sequential or asynchronous) message
scheduling is termed `serial GaBP'.

\subsection{Numerical Example: Toy Linear System: $3\times3$ Equations}
Consider the following $3\times3$ linear system \BE
\underbrace{\left(
                                                 \begin{array}{lll}
                                                   A_{xx}=1 & A_{xy}=-2 & A_{xz}=3 \\
                                                   A_{yx}=-2 & A_{yy}=1 & A_{yz}=0 \\
                                                   A_{zx}=3 & A_{zy}=0 & A_{zz}=1 \\
                                                 \end{array}
                                               \right)}_{\mA}\underbrace{\left(
                                                        \begin{array}{r}
                                                          x \\
                                                          y \\
                                                          z \\
                                                        \end{array}
                                                      \right)}_{\vx}=\underbrace{\left(
                                                                \begin{array}{r}
                                                                  -6 \\
                                                                  0 \\
                                                                  2 \\
                                                                \end{array}
                                                              \right)}_{{\vb}}. \EE
We would like to find the solution to this system,
$\vx^{\ast}=\{x^{\ast},y^{\ast},z^{\ast}\}^{T}$. Inverting the
data matrix $\mA$, we directly solve  \BE \underbrace{\left(
																		\begin{array}{r}
                                                                       x^{\ast} \\
                                                                       y^{\ast} \\
                                                                       z^{\ast} \\
                                                                     \end{array}
                                                                   \right)}_{\vx^{\ast}}=
\underbrace{\left(
                                                 \begin{array}{rrr}
                                                   -1/12 & -1/6 & 1/4 \\
                                                   -1/6 & 2/3 & 1/2 \\
                                                   1/4 & 1/2 & 1/4 \\
                                                 \end{array}
                                               \right)}_{\mA^{-1}}\underbrace{\left(
                                                        \begin{array}{r}
                                                          -6 \\
                                                          0 \\
                                                          2 \\
                                                        \end{array}
                                                      \right)}_{\vb}=\left(
                                                                       \begin{array}{r}
                                                                         1 \\
                                                                         2 \\
                                                                        -1 \\
                                                                       \end{array}
                                                                     \right)
. \EE

\begin{figure}[h!]\label{fig_1}
\begin{center}
    \includegraphics[width=0.4\textwidth]{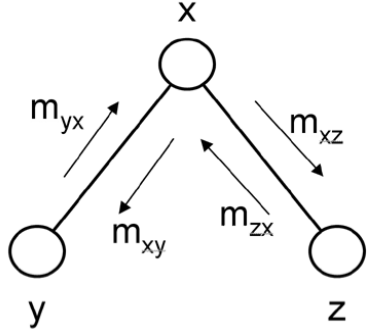}
  \caption{A tree topology with three nodes}
  \label{fig_tree_topo}
\end{center}
\end{figure}

Alternatively, we can now run the GaBP solver.
Fig.~\ref{fig_tree_topo} displays the graph, corresponding to the
data matrix $\mA$, and the message-passing flow. As
$A_{yz}=A_{zy}=0$, this graph is a cycle-free tree, thus GaBP is
guaranteed to converge in a finite number of rounds. As
demonstrated in the following, in this example GaBP converges only
in two rounds, which equals the tree's diameter. Each propagating
message, $m_{ij}$, is described by two scalars $\mu_{ij}$ and
$P_{ij}$, standing for the mean and precision of this
distribution. The evolution of the propagating means and
precisions, until convergence, is described in
Table~\ref{tab_evolve}, where the notation $t=0,1,2,3$ denotes the
iteration rounds. Converged values are written in bold.
\begin{table}[h!]
\begin{center}
$\begin{array}{|l|l|r|r|r|r|}
    \hline \textrm{Message}&\textrm{Computation}&\textrm{t=0}&\textrm{t=1}&\textrm{t=2}&\textrm{t=3}\\\hline\hline
  P_{xy} & -A_{xy}^{2}/(P_{xx}+P_{zx}) & 0 & -4& 1/2 & \mathbf{1/2} \\\hline
  P_{yx} & -A_{yx}^{2}/(P_{yy}) & 0 & -4 & \mathbf{-4} & \mathbf{-4} \\\hline
  P_{xz} & -A_{xz}^{2}/(P_{zz}) & 0 & -9 & 3 & \mathbf{3} \\\hline
  P_{zx} & -A_{zx}^{2}/(P_{xx}+P_{yx}) & 0 & -9 & \mathbf{-9} & \mathbf{-9} \\\hline
  \mu_{xy} & (P_{xx}\mu_{xx}+P_{zx}\mu_{zx})/A_{xy} & 0 & 3 & 6 & \mathbf{6} \\\hline
  \mu_{yx} & P_{yy}\mu_{yy}/A_{yx} & 0 & \mathbf{0} & \mathbf{0} & \mathbf{0} \\\hline
  \mu_{xz} & (P_{xx}\mu_{xx}+P_{yx}\mu_{yx})/A_{xz} & 0 & -2 & \mathbf{-2} & \mathbf{-2} \\\hline
  \mu_{zx} & P_{zz}\mu_{zz}/A_{zx} & 0 & 2/3 & \mathbf{2/3} & \mathbf{2/3}\\\hline
\end{array}$
\label{tab_evolve}
\end{center}
\caption{Evolution of means and precisions on a tree with three
nodes}
\end{table}

Next, following the GaBP solver algorithm, we infer the marginal
means. For exposition purposes we also present in
Table~\ref{tentative_means} the tentative solutions at each
iteration round.
\begin{table}[h!]
\begin{center}
$
\begin{array}{|l|l|r|r|r|r|}
    \hline \textrm{Solution}&\textrm{Computation}&\textrm{t=0}&\textrm{t=1}&\textrm{t=2}&\textrm{t=3}\\\hline\hline
  \mu_{x} & \big(P_{xx}\mu_{xx}+P_{zx}\mu_{zx}+P_{yx}\mu_{yx}\big)/\big(P_{xx}+P_{zx}+P_{yx}\big) & -6 & 1 & \mathbf{1} & \mathbf{1} \\\hline
  \mu_{y} & \big(P_{yy}\mu_{yy}+P_{xy}\mu_{xy}\big)/\big(P_{yy}+P_{xy}\big) & 0 & 4 & 2 & \mathbf{2}\\\hline
  \mu_{z} & \big(P_{zz}\mu_{zz}+P_{xz}\mu_{xz}\big)/\big(P_{zz}+P_{xz}\big) & 2 & -5/2& -1 & \mathbf{-1}\\\hline
\end{array}.
$ \comment{ \BEA
\mu_{x}&=&\big(P_{xx}\mu_{xx}+P_{zx}\mu_{zx}+P_{yx}\mu_{yx}\big)/\big(P_{xx}+P_{zx}+P_{yx}\big)=1,\\
\mu_{y}&=&\big(P_{yy}\mu_{yy}+P_{xy}\mu_{xy}\big)/\big(P_{yy}+P_{xy}\big)=2,\\
\mu_{z}&=&\big(P_{zz}\mu_{zz}+P_{xz}\mu_{xz}\big)/\big(P_{zz}+P_{xz}\big)=-1,
\EEA}
\end{center}
\caption{Tentative means computed on each iteration until
convergence} \label{tentative_means}
\end{table}

Thus, as expected, the GaBP solution
$\vx^{\ast}=\{x^{\ast}=1,y^{\ast}=2,z^{\ast}=-1\}^{T}$ is
identical to what is found taking the direct approach. Note that
as the linear system is described by a tree graph, then for this
particular case, the inferred precision is also exact \BEA
P_{x}&=&P_{xx}+P_{yx}+P_{zx}=-12,\\
P_{y}&=&P_{yy}+P_{xy}=3/2,\\
P_{z}&=&P_{zz}+P_{xz}=4.\\
\EEA and gives
$\{P_{x}^{-1}=\{\mA^{-1}\}_{xx}=-1/12,P_{y}^{-1}=\{\mA^{-1}\}_{yy}=2/3,P_{z}^{-1}=\{\mA^{-1}\}_{zz}=1/4\}^{T}$,
\ie the true diagonal values of the data matrix's inverse,
$\mA^{-1}$.

\subsection{Application Example: Linear Detection}\label{sec_linear}

Consider a discrete-time channel with a real input vector
\mbox{$\vx=\{x_{1},\ldots,x_{K}\}^{T}$} governed by an arbitrary prior
distribution, $P_{\vx}$, and a corresponding real
output vector
$\vy=\{y_{1},\ldots,y_{K}\}^{T}=f\{\vx^{T}\}\in\mathbb{R}^{K}$.\footnote{An
extension to the complex domain is straightforward.} Here, the
function $f\{\cdot\}$ denotes the channel transformation. By
definition, linear detection compels the decision rule to
be\BE\label{eq_gld}
\hat{\vx}=\Delta\{\vx^{\ast}\}=\Delta\{\mA^{-1}\vb\}, \EE where
$\vb=\vy$ is the $K\times 1$ observation vector and the $K\times
K$ matrix $\mA$ is a positive-definite symmetric matrix
approximating the channel transformation. The vector $\vx^{\ast}$
is the solution (over $\mathbb{R}$) to $\mA\vx=\vb$. Estimation is
completed by adjusting the (inverse) matrix-vector product to the
input alphabet, dictated by $P_{\vx}$, accomplished by using a
proper clipping function $\Delta\{\cdot\}$ (\eg, for binary
signaling $\Delta\{\cdot\}$ is the sign function).

For example, linear channels, which appear extensively in many
applications in communication and data storage systems, are
characterized by the linear relation\BE \vy=f\{\vx\}=\mR\vx+\vn,
\EE where $\vn$ is a $K\times 1$ additive noise vector and
\mbox{$\mR=\mS^{T}\mS$} is a positive-definite symmetric matrix,
often known as the correlation matrix. The $N\times K$ matrix
$\mS$ describes the physical channel medium while the vector $\vy$
corresponds to the output of a bank of filters matched to the
physical channel $\mS$.

Due to the vast applicability of linear channels, in
Section~\ref{sec_results} we focus in our experimental study on
such channels, although our paradigm is not limited to this case.
Assuming linear channels with AWGN with variance $\sigma^{2}$ as
the ambient noise, the general linear detection
rule~(\ref{eq_gld}) can describe known linear detectors.
For example ~\cite{BibDB:BookVerdu,BibDB:BookProakis}:
\begin{itemize}
                 \item The conventional matched filter (MF) detector is obtained by taking \mbox{$\mA\triangleq\mI_{K}$} and \mbox{$\vb=\vy$}. This detector is optimal, in the MAP-sense, for the case of zero cross-correlations, \ie, $\mR=\mI_{K}$, as happens for orthogonal CDMA or when there is no ISI effect.
\item The decorrelator (zero forcing equalizer) is achieved by
substituting \mbox{$\mA\triangleq\mR$} and $\vb=\vy$. It is
optimal in the noiseless case.
                 \item The linear minimum mean-square error (MMSE) detector can also be described by using \mbox{$\mA=\mR+\sigma^{2}\mI_{K}$} and $\vb=\vy$. This detector is 
known to be optimal when the input distribution $P_{\vx}$ is
Gaussian.
               \end{itemize}

In general, linear detection is suboptimal because of its
deterministic underlying mechanism (\ie, solving a given set of
linear equations), in contrast to other estimation schemes, such
as MAP or maximum likelihood, that emerge from an optimization
criterion. In the following section we implement the linear
detection operation, in its general form~(\ref{eq_gld}), in an
efficient message-passing fashion.


The essence of detection theory is to estimate a hidden input to a
channel from empirically-observed outputs. An important class of
practical sub-optimal detectors is based on linear detection. This
class includes, for instance, the conventional single-user matched
filter, the decorrelator (also, called the zero-forcing
equalizer), the linear minimum mean-square error (MMSE) detector,
and many other detectors with widespread
applicability~\cite{BibDB:BookVerdu,BibDB:BookProakis}. In general
terms, given a probabilistic estimation problem, linear detection
solves a deterministic system of linear equations derived from the
original problem, thereby providing a sub-optimal, but often
useful, estimate of the unknown input.

Applying the GaBP solver to linear detection, we establish a new
and explicit link between BP and linear detection. This link
strengthens the connection between message-passing inference and
estimation theory, previously seen in the context of optimal
maximum a-posteriori (MAP)
detection~\cite{BibDB:Kabashima,BibDB:ShentalITW} and several
sub-optimal nonlinear detection
techniques~\cite{BibDB:TanakaOkada} applied in the context of both
dense and sparse~\cite{BibDB:MontanariTse,BibDB:ConfWangGuo}
graphical models.

In the following experimental study, we examine the implementation
of a decorrelator detector in a noiseless synchronous CDMA system
with binary signaling and spreading codes based upon Gold
sequences of length $m=7$.\footnote{In this case, as long as the
system is not overloaded, \ie the number of active users $n$ is
not greater than the spreading code's length $m$, the decorrelator
detector yields optimal detection decisions.} Two system setups
are simulated,  corresponding to $n=3$ and $n=4$ users, resulting
in the cross-correlation matrices  \BE
\mR_{3} = \frac{1}{7}\left(%
\begin{array}{rrr}
  7 & -1 & 3 \\
  -1 & 7 & -5 \\
  3 & -5 & 7 \\
\end{array} \right) \EE and
\BE \mR_{4} = \frac{1}{7}\left(%
\begin{array}{rrrr}
  7 & -1 & 3 & 3\\
  -1 & 7 & 3 & -1\\
  3 & 3 & 7 & -1\\
  3 & -1 & -1 & 7\\
\end{array} \right), \EE respectively.\footnote{These particular correlation settings were taken from the simulation setup of Yener \etal~\cite{BibDB:YenerEtAl}.}

The decorrelator detector, a member of the family of linear
detectors, solves a system of linear equations,
\mbox{$\mA\vx=\vb$}, where the matrix $\mA$ is equal to the
$n\times n$ correlation matrix $\mR$, and the observation vector
$\vb$ is identical to the $n$-length CDMA channel output vector
$\vy$.  Thus the vector of decorrelator decisions is determined by
taking the signum of the vector $\mA^{-1}\vb = \mR^{-1}\vy$. Note
that $\mR_{3}$ and $\mR_{4}$ are not strictly diagonally dominant,
but their spectral radii are less than unity, since
$\rho(|\mI_{3}-\mR_{3}|)=0.9008<1$ and
$\rho(|\mI_{4}-\mR_{4}|)=0.8747<1$, respectively. In all of the
experiments, we assumed the output vector was the all-ones vector.

Table~\ref{tab_1} compares the proposed GaBP algorithm with
standard iterative solution methods~\cite{BibDB:BookAxelsson}
(using random initial guesses), previously employed for CDMA
multiuser detectors (MUD). Specifically, MUD algorithms based on
the algorithms of Jacobi, Gauss-Seidel (GS) and (optimally
weighted) successive over-relaxation (SOR)\footnote{This moving
average improvement of Jacobi and GS algorithms is equivalent to
what is known in the BP literature as `damping'~\cite{Damping}.}
were investigated~\cite{grant99iterative,BibDB:TanRasmussen}. The
table lists the convergence rates for the two Gold code-based CDMA
settings. Convergence is identified and declared when the
differences in all the iterated values are less than $10^{-6}$. We
see that, in comparison with the previously proposed detectors
based upon the Jacobi and GS algorithms, the GaBP detectors
converge more rapidly for both $n=3$ and $n=4$. The serial
(asynchronous) GaBP algorithm achieves the best overall
convergence rate, surpassing even the SOR-based detector. 

\begin{figure}[h!]
\begin{center}
    \includegraphics[width=0.35\textwidth]{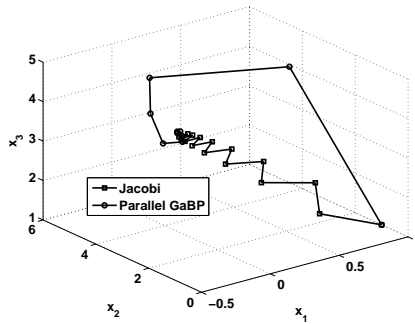}
  \caption{Convergence of the GaBP algorithm vs. Jacobi on a $3 \times 3$ gold CDMA matrix. Each dimension shows
  one coordinate of the solution. Jacobi converges in zigzags while GaBP has spiral convergence.}
\label{fig_R3_spiral}
\end{center}
\end{figure}

\begin{figure}[t!]
\begin{minipage}[b]{0.5\linewidth}
\centering
    \includegraphics[width=\textwidth]{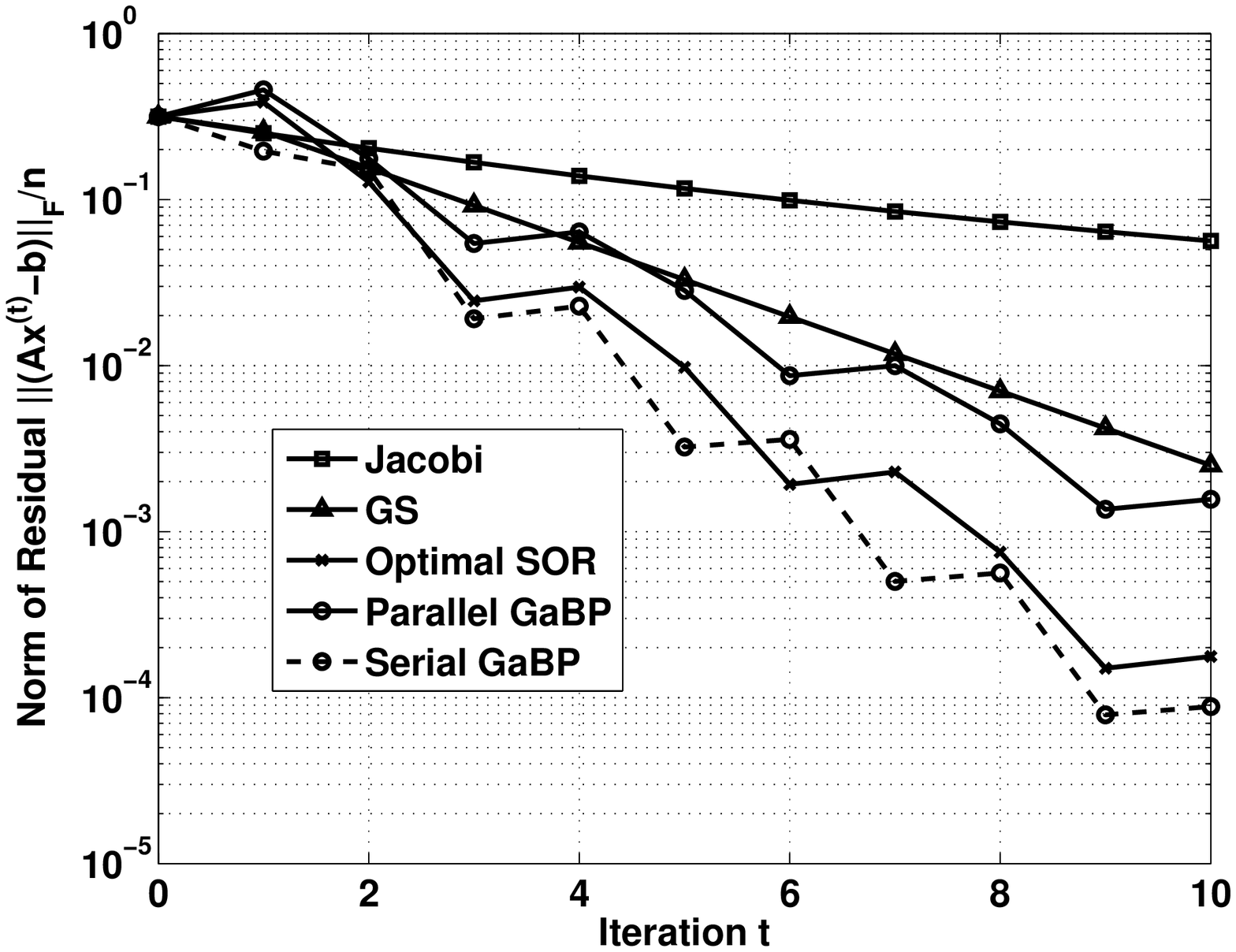}
 \label{fig_R3}
\end{minipage}
\begin{minipage}[b]{0.5\linewidth}
\centering
   \includegraphics[width=\textwidth]{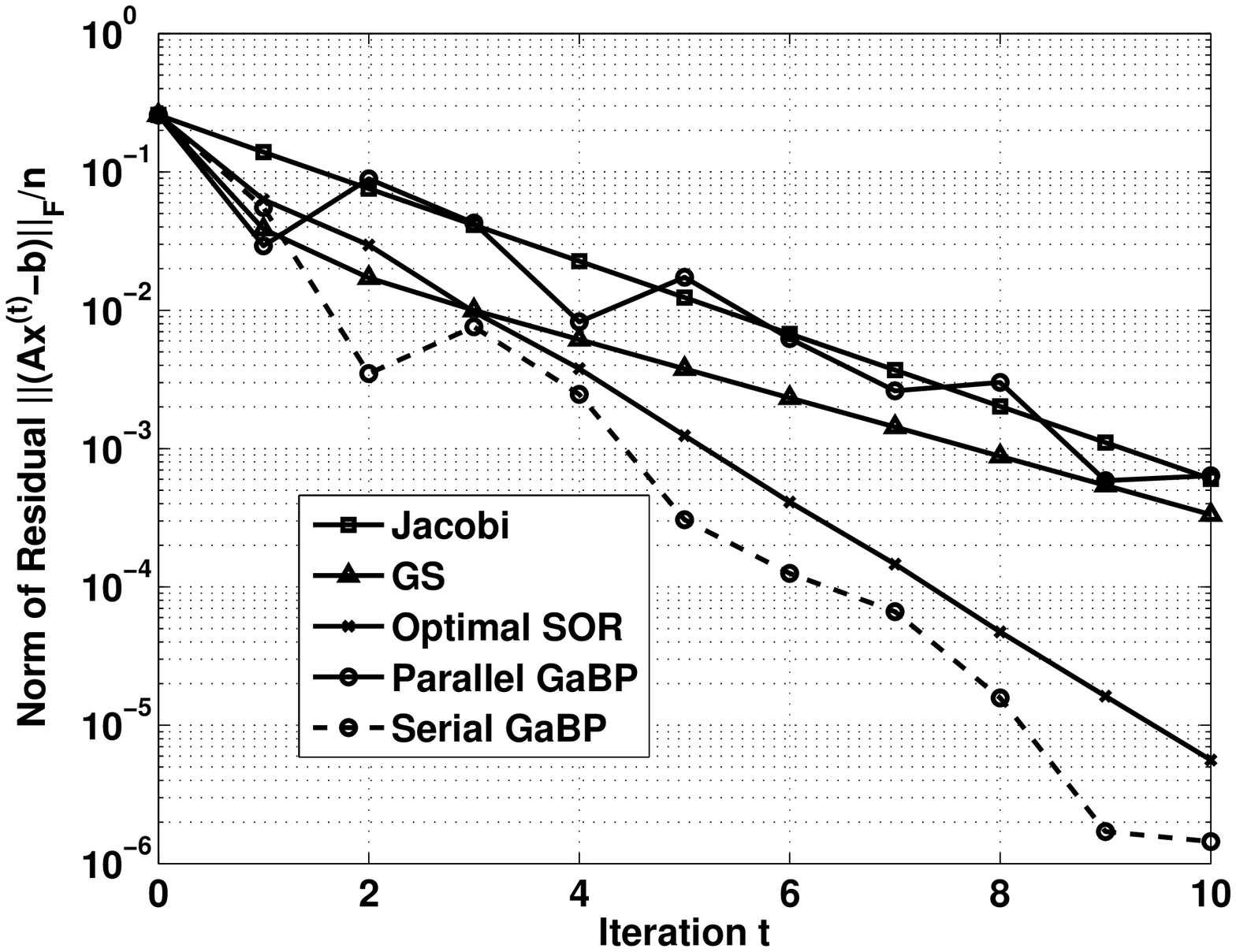}
 \label{fig_R4}
\end{minipage}
\caption{Convergence of the two gold CDMA matrices. To the left $\mR_3$, to the right, $\mR_4$. }
\end{figure}

\begin{table}
\centerline{
\begin{tabular}{|c|r|r|}
  \hline
  \textbf{Algorithm}
  & Iterations $t$ ($\mR_{3}$0 & Iterations $t$ ($\mR_{4}$) \\\hline\hline & & \\
  Jacobi & 111\comment{without dividing R_{3} with 7: 136} & 24\comment{50} \\\hline & & \\
  GS & 26 & 26\comment{32}\\\hline & & \\
  \textbf{Parallel GaBP} & \textbf{23} & \textbf{24}\\\hline & & \\
  Optimal SOR & 17 & 14\comment{20} \\\hline & & \\
  \textbf{Serial GaBP} & \textbf{16} & \textbf{13}\\
  \hline
\end{tabular}
}\vspace{0.5cm} \caption{Decorrelator for $K=3,4$-user, $N=7$ Gold
code CDMA. Total number of iterations required for convergence
(threshold $\epsilon=10^{-6}$) for GaBP-based solvers vs. standard
methods.}\label{tab_1}
\end{table}

\begin{table}[h!]
\centerline{
\begin{tabular}{|l|c|c|}
  \hline
  \textbf{Algorithm}
  & $\mR_{3}$ & $\mR_{4}$ \\
  \hline\hline & & \\
  Jacobi+Steffensen\footnotemark & 59\comment{51} & $-$ \\\hline & & \\
  \textbf{Parallel GaBP+Steffensen} & \textbf{13} & \textbf{13}\\\hline & & \\
  \textbf{Serial GaBP+Steffensen} & \textbf{9} & \textbf{7} \\
  \hline
\end{tabular}
}\vspace{0.5cm}\caption{Decorrelator for $K=3,4$-user, $N=7$ Gold
code CDMA. Total number of iterations required for convergence
(threshold $\epsilon=10^{-6}$) for Jacobi, parallel and serial
GaBP solvers accelerated by Steffensen iterations.}\label{tab_2}
\end{table}

Further speed-up of GaBP can be achieved by adapting known
acceleration techniques from linear algebra, such Aitken's method and Steffensen's iterations~\cite{BibDB:BookHenrici}. Consider a sequence $\{x_{n}\}$ (\eg, obtained by using
GaBP iterations) linearly converging to the limit $\hat{x}$, and
$x_n \ne \hat{x}$ for $n \ge 0$. According to Aitken's method, if
there exists a real number $a$ such that $|a|<1 $ and
\mbox{$\lim_{n \rightarrow \infty}(x_n-\hat{x})/(x_{n-1} -
\hat{x}) = a$}, then the sequence $\{ y_n\}$ defined by
\[ y_n = x_n - \frac{(x_{n+1} -x_n)^2}{x_{n+2} - 2x_{n+1} + x_n} \]
converges to $\hat{x}$ faster than $\{ x_n \}$ in the sense that
\mbox{$\lim_{n \rightarrow \infty} |(\hat{x} - y_n)/(\hat{x} -
x_n)| = 0$}. Aitken's method can be viewed as a generalization of
over-relaxation, since one uses values from three, rather than
two, consecutive iteration rounds. This method can be easily
implemented in GaBP as every node computes values based only on its own
history.

\begin{figure}[t!]
\begin{minipage}[b]{0.5\linewidth}
\centering
     \includegraphics[width=\textwidth]{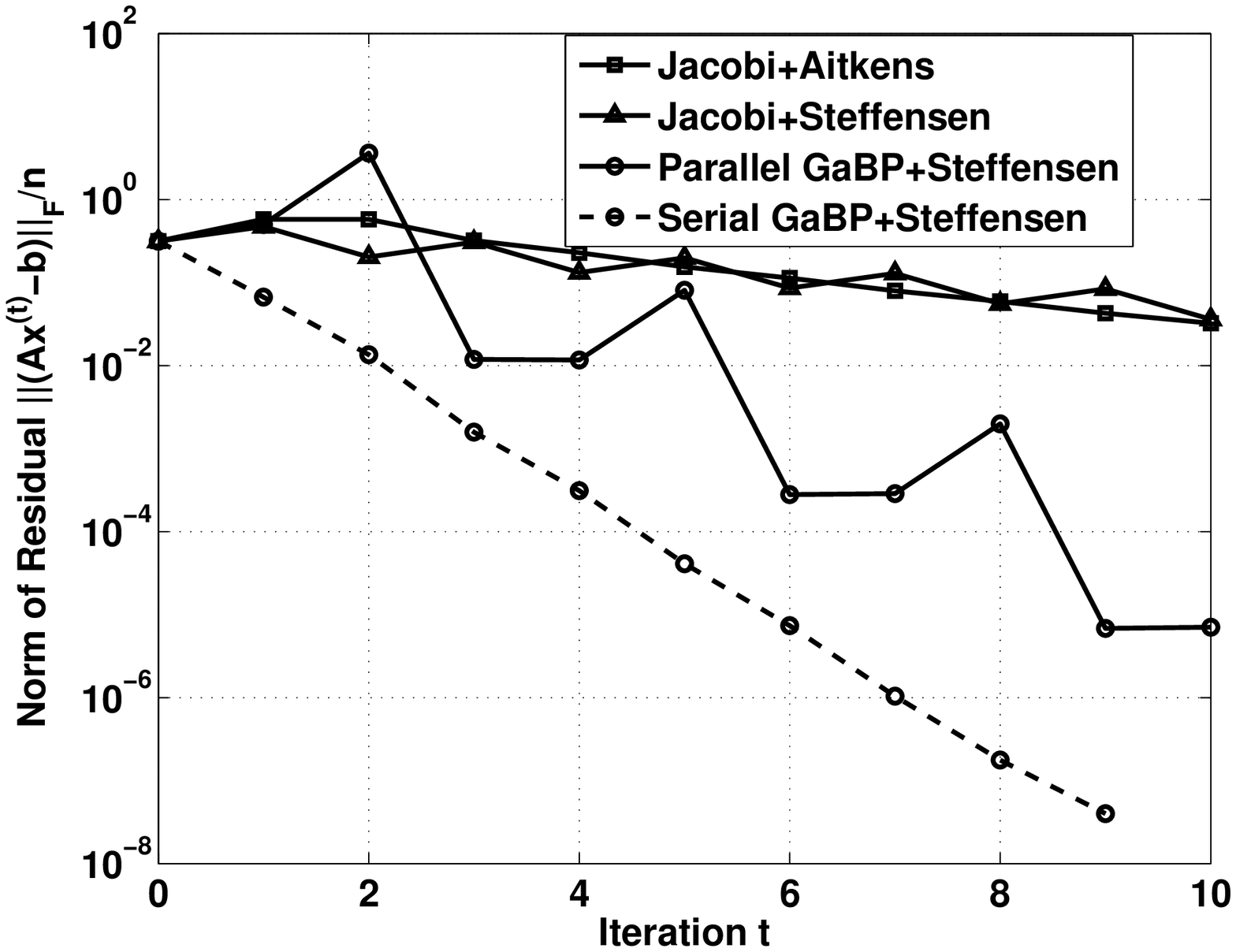}
  \label{fig_R3_accel}
\end{minipage}
\begin{minipage}[b]{0.5\linewidth}
\centering
\vspace{5mm}
   \includegraphics[width=\textwidth]{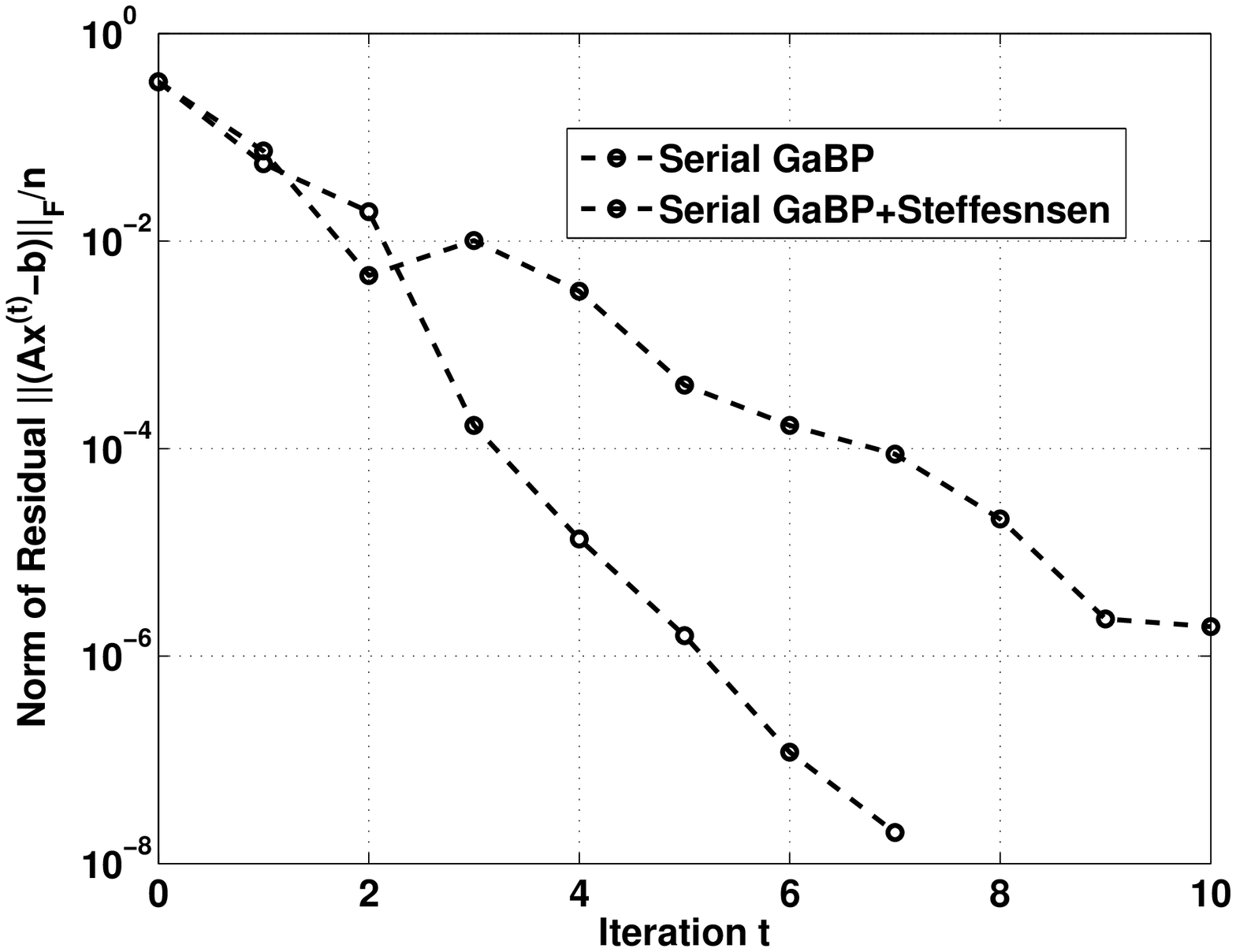}
  \label{fig_R4_accel}
\end{minipage}
 \caption{Convergence acceleration of the GaBP algorithm using Aitken and Steffensen methods.
  The left graph depicts a $3 \times 3$ gold CDMA matrix, the right graph $4 \times 4$ gold
  CDMA matrix. Further details regarding simulation setup are found in Section \ref{sec_linear}.}
\end{figure}

Steffensen's iterations incorporate Aitken's method. Starting with
$x_{n}$, two iterations are run to get $x_{n+1}$ and $x_{n+2}$.
Next, Aitken's method is used to compute $y_{n}$, this value
replaces the original $x_{n}$, and GaBP is executed again to
get a new value of $x_{n+1}$. This process is repeated iteratively
until convergence. Table~\ref{tab_2} demonstrates the speed-up of
GaBP obtained by using these acceleration methods, in comparison with that achieved by the similarly modified Jacobi algorithm.\footnote{Application of Aitken and
Steffensen's methods for speeding-up the convergence of standard
(non-BP) iterative solution algorithms in the context of MUD was
introduced by Leibig \etal~\cite{LDF05}.} We remark that, although the convergence rate is improved with these enhanced algorithms, the region of convergence of the accelerated GaBP solver remains unchanged.

For the algorithms we examined, Fig.~1-(a) displays the Euclidean distance between the tentative (intermediate) results and the fixed-point solution
as a function of the number of iterations. As expected,
all linear algorithms exhibit a logarithmic convergence behaviour.
Note that GaBP converges faster on average, although there are some fluctuations in the GaBP curves, in contrast to the monotonicity of the other curves.

An interesting question concerns the origin of this convergence
speed-up associated with GaBP. Better understanding may be gained
by visualizing the iterations of the different methods for the
matrix $\mR_{3}$ case. The convergence contours are plotted in the
space of $\{x_{1},x_{2},x_{3}\}$ in Fig.~1-(b). As expected, the
Jacobi algorithm converges in zigzags towards the fixed point
(this behavior is well-explained in Bertsekas and
Tsitsiklis~\cite{BibDB:BookBertsekasTsitsiklis}).\comment{SOR
algorithm averages the Jacobi algorithm convergence behavior thus
gains faster convergence speed. Aitken's speeds-up Jacobi's
convergence.} The fastest algorithm is serial GaBP. It is
interesting to note that GaBP convergence is in a spiral shape,
hinting that despite the overall convergence improvement,
performance improvement is not guaranteed in successive iteration
rounds. The spiral nature of GaBP convergence is better viewed in
Fig.~1-(c). In this case the system was simulated with a specific $\mR$ matrix for which Jacobi algorithm and other standard methods did not even converge. Using Aitken's method, a further speed-up in GaBP convergence was obtained.

Despite the fact that the examples considered correspond to small multi-user systems, we believe that the results reflect the typical behavior of the algorithms, and that similar qualitative results would be observed in larger systems. In support of this belief, we note, in passing, that GaBP was experimentally shown to converge in a logarithmic number of iterations in the
cases of very large matrices both dense (with up to hundreds of
thousands of dimensions~\cite{ECCS08}) and sparse (with up to
millions of dimensions~\cite{Rating,PPNA08}).

As a final remark on the linear detection example, we mention that, in the case of a channel with Gaussian input signals,
for which linear detection is optimal, it can be easily shown that
the proposed GaBP scheme reduces to the BP-based MUD scheme,
recently introduced by Montanari \etal~\cite{BibDB:MontanariEtAl}.
Their BP scheme, tailored specifically for Gaussian signaling, has
been proven to converge to the MMSE (and optimal) solution for any
arbitrarily loaded, randomly-spread CDMA system (\ie, a system
where $\rho(\mI_{n}-\mR)\lesseqgtr1$). \footnote{For non-Gaussian
signaling, \eg with binary input alphabet, this BP-based detector
is conjectured to converge only in the large-system limit, as
$n,m\rightarrow\infty$~\cite{BibDB:MontanariEtAl}.} Thus
Gaussian-input additive white Gaussian noise CDMA is another
example for which the proposed GaBP solver converges to the MAP
decisions for any $m\times n$ random spreading matrix $\mS$,
regardless of the spectral radius.

\subsection{Numerical Example: Symmetric Non-PSD Indefinite) Data
Matrix}\label{sec_nonPSD}

Consider the case of a linear system with a symmetric, but non-PSD
data matrix \BE
\left(%
\begin{array}{ccc}
  1 & 2 & 3 \\
  2 & 2 & 1 \\
  3 & 1 & 1 \\
\end{array}%
\right). \EE

Table~\ref{tab_nonPSD} displays the number of iterations required
for convergence for the iterative methods under consideration. The
classical methods diverge, even when aided with acceleration
techniques. This behavior (at least without the acceleration) is
not surprising in light of Theorem~\ref{spectral_radius_thm}. Again we observe that
serial scheduling of the GaBP solver is superior parallel
scheduling and that applying Steffensen iterations reduces the
number of iterations in $45\%$ in both cases. Note that SOR cannot
be defined when the matrix is not PSD. By definition CG works only
for symmetric PSD matrices. because the solution is a saddle point
and not a minimum or maximum.

\begin{table}[h!]
\centerline{
\begin{tabular}{|c|r|}
  \hline
  \textbf{Algorithm}
  & \textbf{Iterations} $t$\\
  \hline\hline & \\
  Jacobi,GS,SR,Jacobi+Aitkens,Jacobi+Steffensen & $-$\\\hline & \\
  \textbf{Parallel GaBP} & \textbf{38} \\\hline & \\
  \textbf{Serial GaBP} & \textbf{25} \\\hline & \\
  \textbf{Parallel GaBP+Steffensen} & \textbf{21} \\\hline & \\
  \textbf{Serial GaBP+Steffensen} & \textbf{14} \\
  \hline
\end{tabular}
}\vspace{0.5cm}\caption{Symmetric non-PSD $3\times3$ data matrix.
Total number of iterations required for convergence (threshold
$\epsilon=10^{-6}$) for GaBP-based solvers vs. standard
methods.}\label{tab_nonPSD}
\end{table}

\begin{figure}[h!]\label{fig_S}
\begin{center}
    \includegraphics[width=0.5\textwidth]{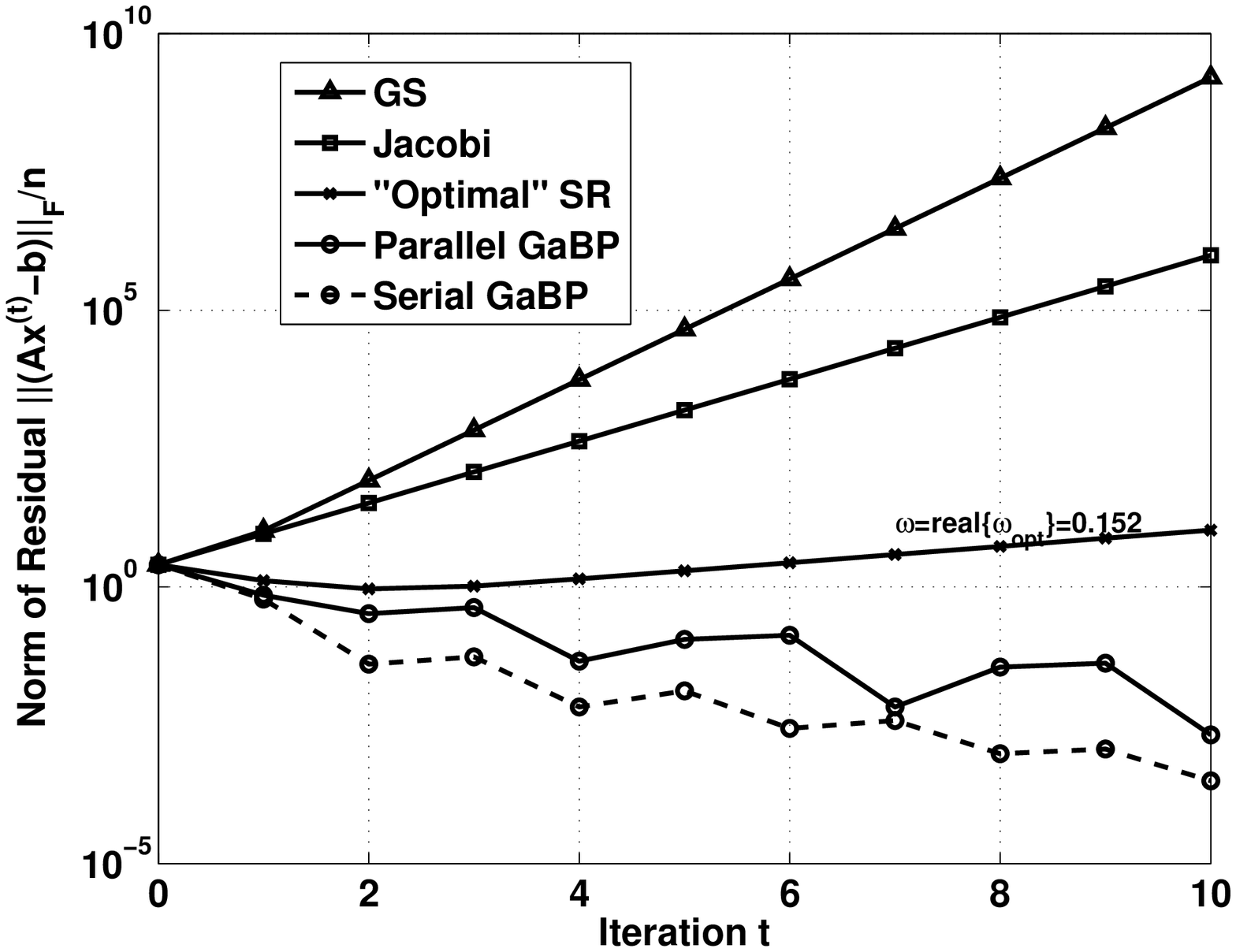}
  \caption{Convergence rate for a $3\times3$ symmetric non-PSD data matrix.
  The Frobenius norm of the residual per equation, $||\mA\vx^{t}-b||_{F}/n$, as a function of the iteration $t$
  for GS (triangles and solid line), Jacobi (squares and solid line), SR (stars and solid line), parallel GaBP (circles and solid line)
  and serial GaBP (circles and dashed line) solvers.}
\end{center}
\end{figure}
\begin{figure}[t!]
\begin{minipage}[b]{0.5\linewidth}
\centering
     \includegraphics[width=\textwidth]{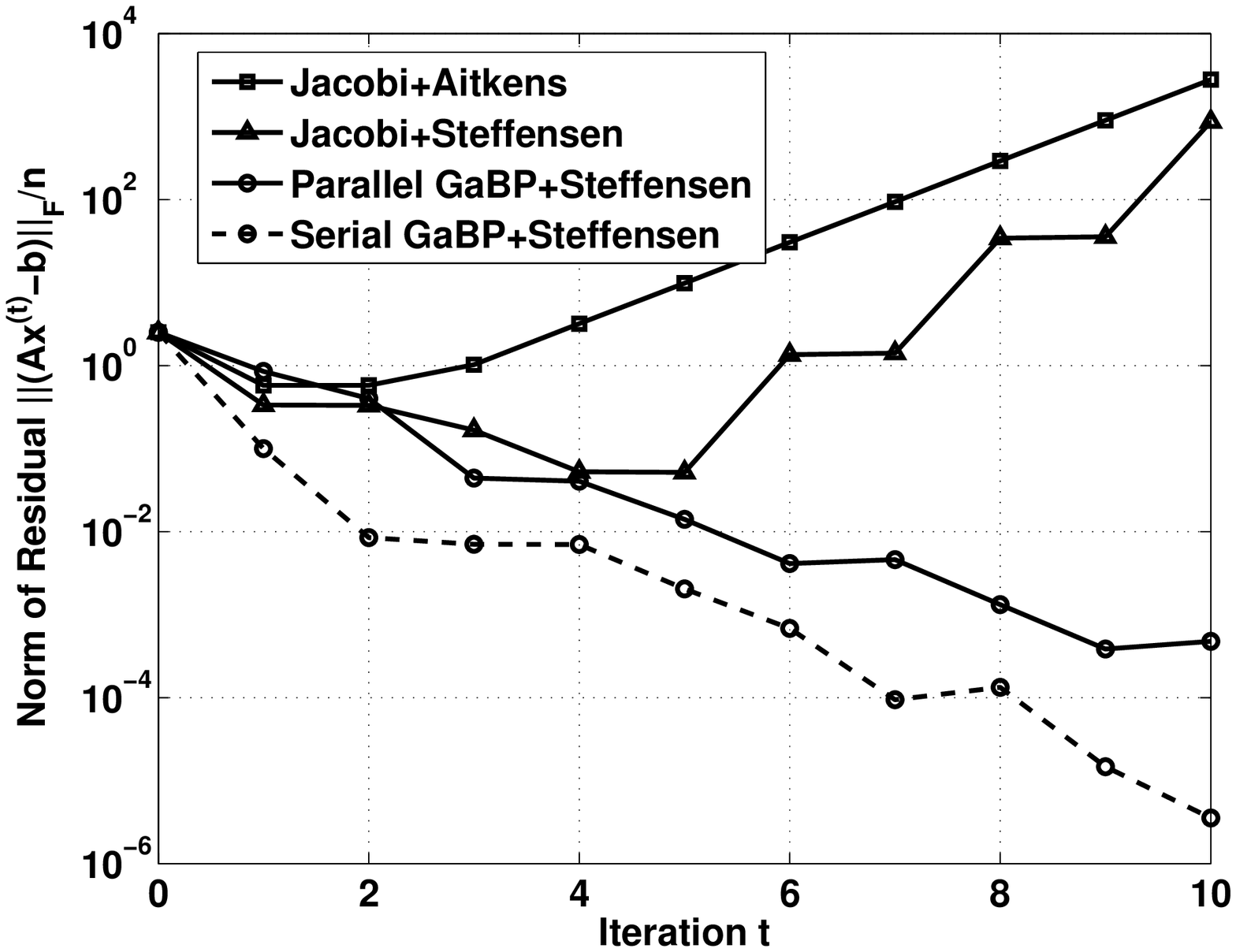}
 \label{fig_S_accel}
\end{minipage}
\begin{minipage}[b]{0.5\linewidth}
\centering
   \includegraphics[width=\textwidth]{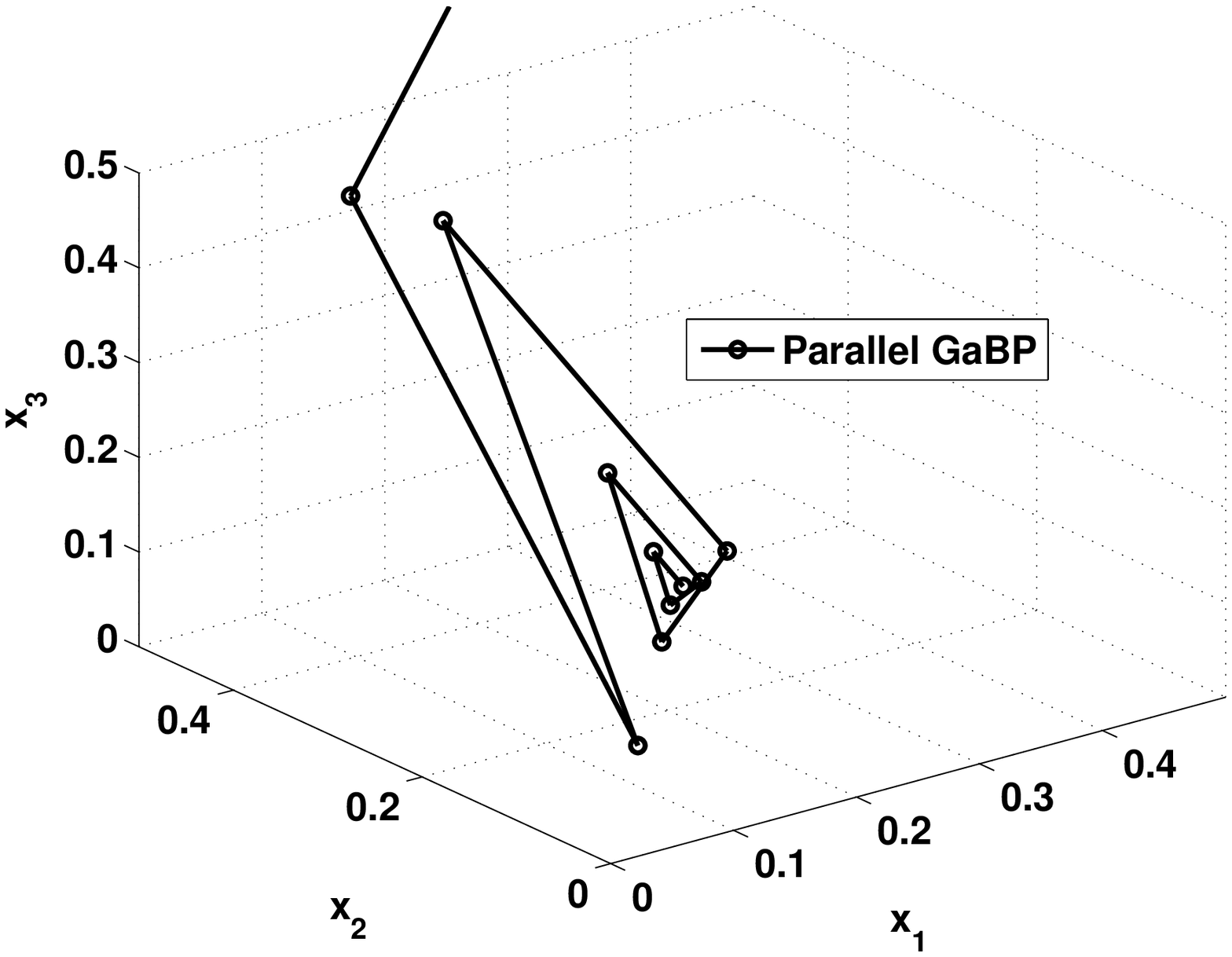}
 \label{fig_S_spiral}
\end{minipage}
\caption{The left graph depicts accelerated convergence rate for a $3\times3$ symmetric non-PSD data matrix.
  The Frobenius norm of the residual per equation, $||\mA\vx^{t}-b||_{F}/n$, as a function of the iteration $t$
  for Aitkens (squares and solid line) and Steffensen-accelerated (triangles and solid line) Jacobi method, parallel GaBP (circles and solid line)   and serial GaBP (circles and dashed line) solvers accelerated by Steffensen iterations.
  The right graph shows a visualization of parallel GaBP on the same problem, drawn in $\mathbb{R}^{3}$.}
\end{figure}

\section[2D Poisson's]{Application Example: 2-D Poisson's Equation} One of the
most common partial differential equations (PDEs) encountered in
various areas of exact sciences and engineering (\eg, heat flow,
electrostatics, gravity, fluid flow, quantum mechanics,
elasticity) is Poisson's equation. In two dimensions, the equation
is \BE \Delta u(x,y)=f(x,y), \label{eq_poisson} \EE for $\{x,y\}\in\Omega$, where \BE
\Delta{(\cdot)}=\pdpd{(\cdot)}{x}+\pdpd{(\cdot)}{y}. \EE is the
Laplacean operator and $\Omega$ is a bounded domain in
$\mathbb{R}^{2}$. The solution is well defined only under boundary
conditions, \ie, the value of $u(x,y)$ on the boundary of $\Omega$
is specified. We consider the simple (Dirichlet) case of
$u(x,y)=0$ for \{x,y\} on the boundary of $\Omega$. This equation
describes, for instance, the steady-state temperature of a uniform
square plate with the boundaries held at temperature $u=0$, and
$f(x,y)$ equaling the external heat supplied at point $\{x,y\}$.

\begin{figure}[h!]\label{fig_P_accel}
\begin{center}
    \includegraphics[width=0.5\textwidth]{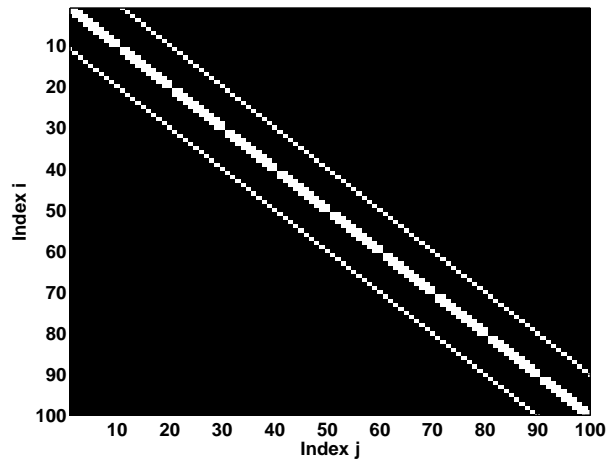}
  \caption{Image of the corresponding sparse data matrix for the 2-D discrete Poisson's PDE with $p=10$. Empty (full) squares denote non-zero (zero) entries.}
  \label{2D_Poisson}
\end{center}
\end{figure}

The poisson's PDE can be discretized by using finite differences.
An $p+1\times p+1$ square grid on $\Omega$ with size (arbitrarily)
set to unity is used, where $h\triangleq1/(p+1)$ is the grid
spacing. We let $U(i,j)$, $\{i,j=0,\ldots,p+1\}$, be the
approximate solution to the PDE at $x=ih$ and $y=jh$.
Approximating the Laplacean by \BEA \Delta
U(x,y)&=&\pdpd{(U(x,y))}{x}+\pdpd{(U(x,y))}{y}\nonumber\\&\approx&\frac{U(i+1,j)-2U(i,j)+U(i-1,j)}{h^{2}}+\frac{U(i,j+1)-2U(i,j)+U(i,j-1)}{h^{2}}\nonumber\\,
\EEA one gets the system of $n=p^{2}$ linear equations with $n$
unknowns \BE 4U(i,j)-U(i-1,j)-U(i+1,j)-U(i,j-1)-U(i,j+1)=b(i,j)
\forall i,j=1,\ldots,p, \EE where
$b(i,j)\triangleq-f(ih,jh)h^{2}$, the scaled value of the function
$f(x,y)$ at the corresponding grid point $\{i,j\}$. Evidently, the
accuracy of this approximation to the PDE increases with $n$.

\begin{figure}[h!]\label{fig_S}
\begin{center}
    \includegraphics[width=0.5\textwidth]{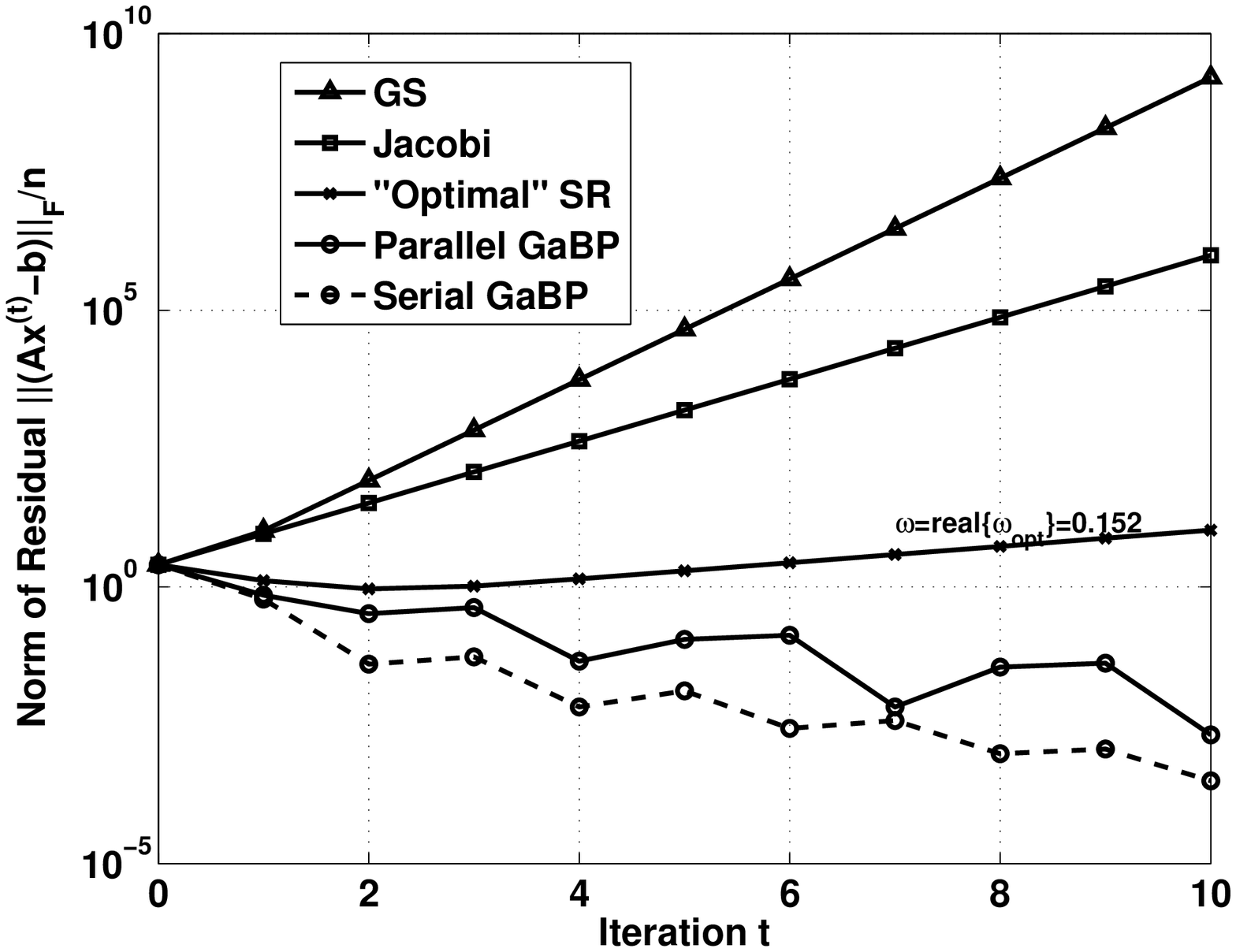}
  \caption{Convergence rate for a $3\times3$ symmetric non-PSD data matrix.
  The Frobenius norm of the residual per equation, $||\mA\vx^{t}-b||_{F}/n$, as a function of the iteration $t$
  for GS (triangles and solid line), Jacobi (squares and solid line), SR (stars and solid line), parallel GaBP (circles and solid line)
  and serial GaBP (circles and dashed line) solvers.}
\end{center}
\end{figure}

\begin{figure}[t!]
\begin{minipage}[b]{0.5\linewidth}
\centering
     \includegraphics[width=\textwidth]{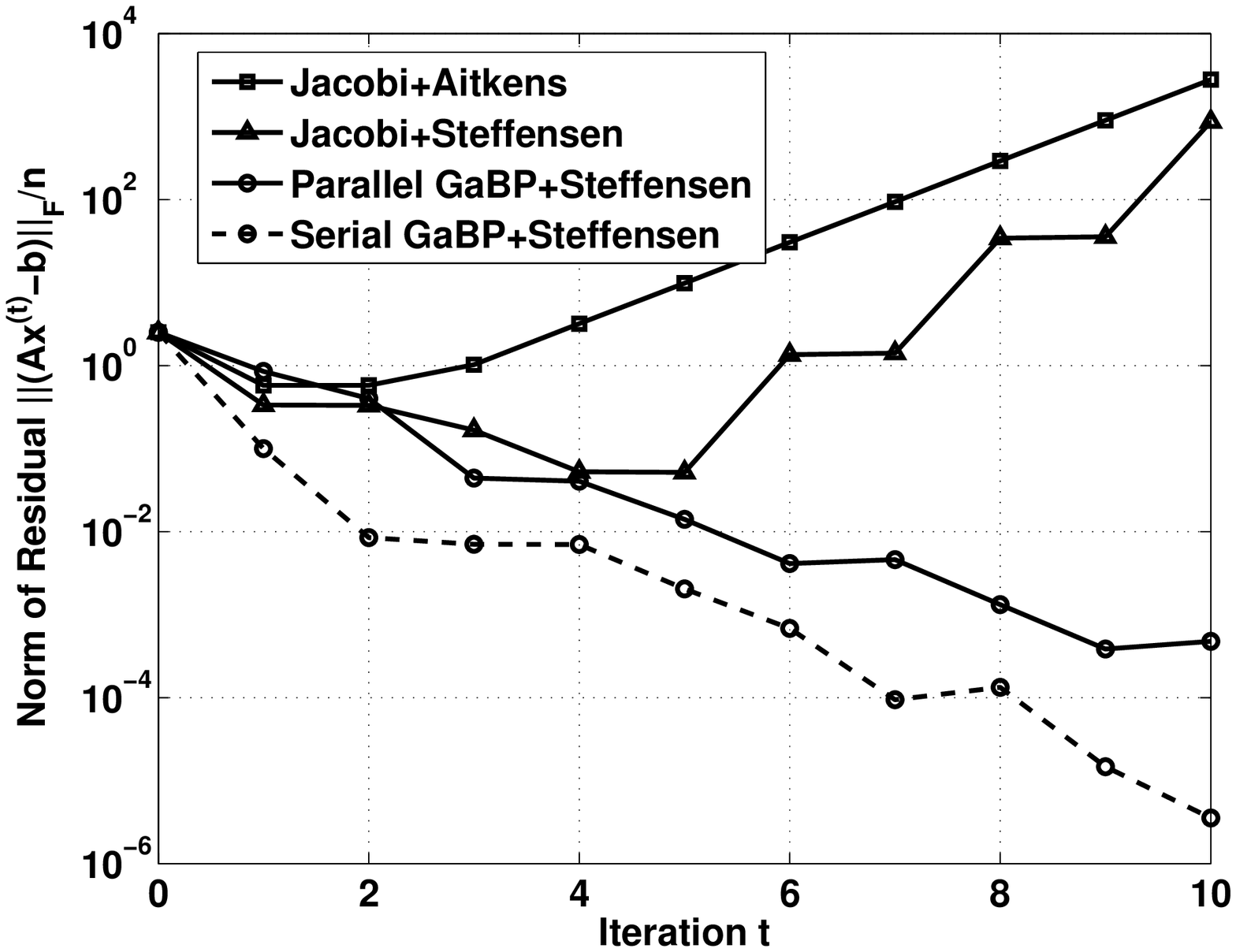}
 \label{fig_S_accel}
\end{minipage}
\begin{minipage}[b]{0.5\linewidth}
\centering
   \includegraphics[width=\textwidth]{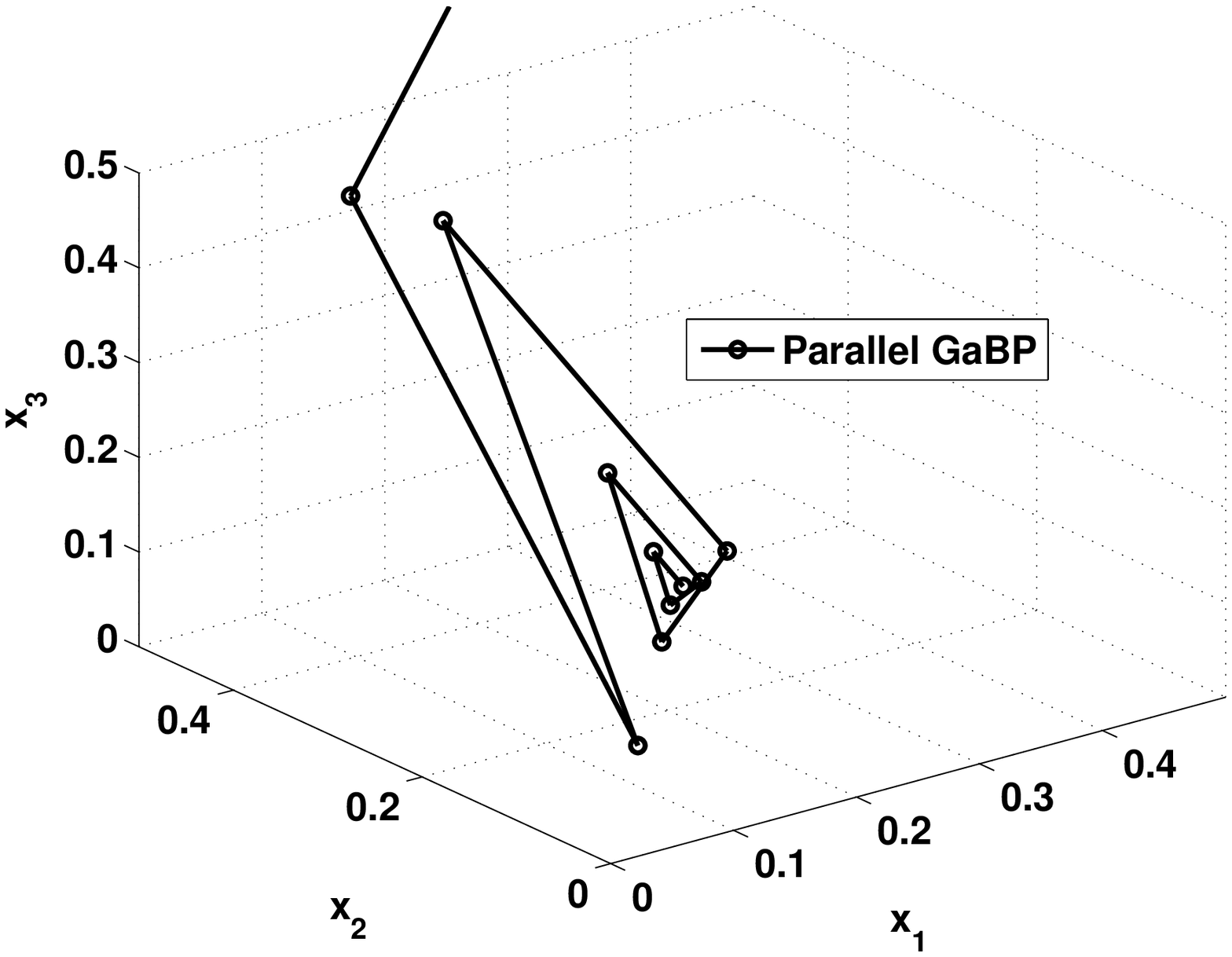}
 \label{fig_S_spiral}
\end{minipage}
\caption{The left graph depicts accelerated convergence rate for a $3\times3$ symmetric non-PSD data matrix.
  The Frobenius norm of the residual per equation, $||\mA\vx^{t}-b||_{F}/n$, as a function of the iteration $t$
  for Aitkens (squares and solid line) and Steffensen-accelerated (triangles and solid line) Jacobi method, parallel GaBP (circles and solid line)   and serial GaBP (circles and dashed line) solvers accelerated by Steffensen iterations.
  The right graph shows a visualization of parallel GaBP on the same problem, drawn in $\mathbb{R}^{3}$.}
\end{figure}

Choosing a certain ordering of the unknowns $U(i,j)$, the linear
system can be written in a matrix-vector form. For example, the
natural row ordering (\ie, enumerating the grid points
left$\rightarrow$right, bottom$\rightarrow$up) leads to a linear
system with $p^{2}\times p^{2}$ sparse data matrix $\mA$. For
example, a Poisson PDE with $p=3$ generates the following
$9\times9$ linear system
\BE\underbrace{\left(%
\begin{array}{ccc|ccc|ccc}
  4 & -1 &  & -1 &  &  &  &\\
  -1 & 4 & -1 &  & -1 &  &  &\\
   & -1 & 4  &  &  & -1 &  &\\\hline
  -1 &  &  & 4 & -1 &  & -1 &\\
   & -1 &  & -1 & 4 & -1 &  & -1\\
   &  & -1 &  & -1 & 4 &  & &-1\\\hline
   &  &  & -1 &  &  & 4 &-1  &\\
   &  &  &  & -1 &  & -1 & 4 &-1\\
   &  &  &  &  & -1 &  & -1 & 4\\
\end{array}%
\right)}_{\mA}\underbrace{\left(%
\begin{array}{c}
  U(1,1) \\ U(2,1) \\ U(3,1) \\ U(1,2) \\ U(2,2) \\ U(3,2)  \\ U(1,3) \\ U(2,3) \\ U(3,3)\\
\end{array}%
\right)}_{\vx}=\underbrace{\left(%
\begin{array}{c}
b(1,1) \\ b(2,1) \\ b(3,1) \\ b(1,2) \\ b(2,2) \\ b(3,2)
\\ b(1,3) \\ b(2,3) \\ b(3,3)\\\end{array}%
\right)}_{\vb},\EE where blank data matrix $\mA$ entries denote
zeros.

\begin{table}
\centerline{
\begin{tabular}{|c|r|}
  \hline
  \textbf{Algorithm}
  & \textbf{Iterations} $t$\\
  \hline\hline & \\
  Jacobi & $354$\\\hline & \\
  GS & $136$\\\hline & \\
  Optimal SOR & $37$\\\hline & \\
  \textbf{Parallel GaBP} & \textbf{134} \\\hline & \\
  \textbf{Serial GaBP} & \textbf{73} \\\hline & \\
  \textbf{Parallel GaBP+Aitkens} & \textbf{25} \\\hline & \\
  \textbf{Parallel GaBP+Steffensen} & \textbf{56} \\\hline & \\
  \textbf{Serial GaBP+Steffensen} & \textbf{32} \\\hline & \\
  \hline
\end{tabular}
}\vspace{0.5cm}\caption{2-D discrete Poisson's PDE with $p=3$ and
$f(x,y)=-1$. Total number of iterations required for convergence
(threshold $\epsilon=10^{-6}$) for GaBP-based solvers vs. standard
methods.}\label{tab_2D_Poisson}
\end{table}

\begin{figure}[h!]\label{fig_P_accel}
\begin{center}
    \includegraphics[width=0.5\textwidth]{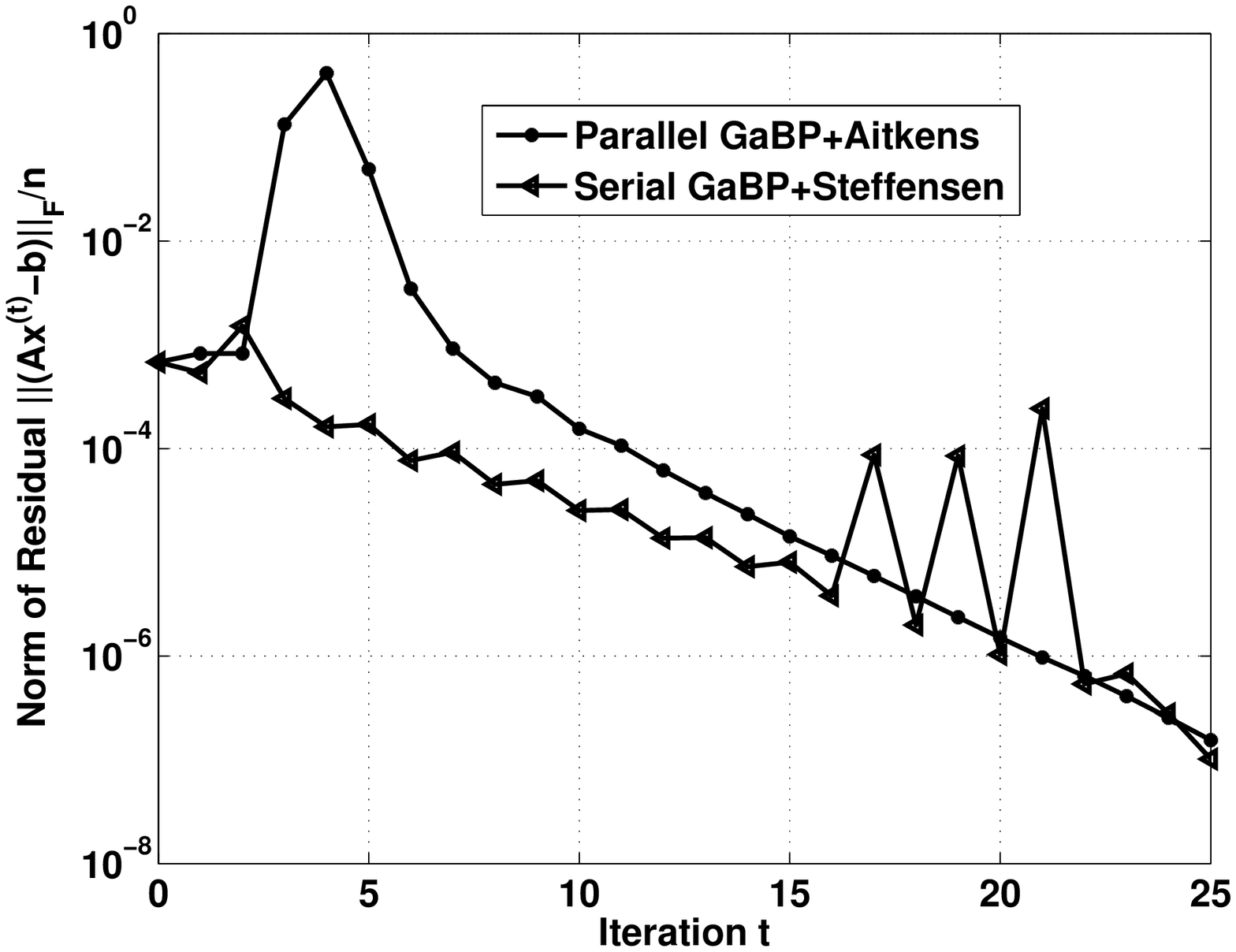}
  \caption{Accelerated convergence rate for the 2-D discrete Poisson's PDE with $p=10$ and $f(x,y)=-1$. The Frobenius norm of the residual. per equation, $||\mA\vx^{t}-b||_{F}/n$, as a function of the iteration $t$ for parallel GaBP solver accelrated by Aitkens method ($\circ$-marks and solid line) and serial GaBP solver accelerated by Steffensen iterations (left triangles and solid line)}.
  \end{center}
\end{figure}

Hence, now we can solve the discretized 2-D Poisson's PDE by
utilizing the GaBP algorithm. Note that, in contrast to the other
examples, in this case the GaBP solver is applied for solving a
sparse, rather than dense, system of linear equations.

In order to evaluate the performance of the GaBP solver, we choose
to solve the 2-D Poisson's equation with discretization of $p=10$.
The structure of the corresponding $100\times100$ sparse data
matrix is illustrated in Fig.~\ref{2D_Poisson}.

\begin{table}
\centerline{
\begin{tabular}{|c|r|}
  \hline
  \textbf{Algorithm}
  & \textbf{Iterations} $t$\\
  \hline\hline & \\
  Jacobi,GS,SR,Jacobi+Aitkens,Jacobi+Steffensen & $-$\\\hline & \\
  \textbf{Parallel GaBP} & \textbf{84} \\\hline & \\
  \textbf{Serial GaBP} & \textbf{30} \\\hline & \\
  \textbf{Parallel GaBP+Steffensen} & \textbf{43} \\\hline & \\
  \textbf{Serial GaBP+Steffensen} & \textbf{17} \\
  \hline
\end{tabular}
}\vspace{0.5cm}\caption{Asymmetric $3\times3$ data matrix. total
number of iterations required for convergence (threshold
$\epsilon=10^{-6}$) for GaBP-based solvers vs. standard
methods.}\label{tab_Asym}
\end{table}

\begin{figure}[t!]
\begin{minipage}[b]{0.5\linewidth}
\centering
    \includegraphics[width=\textwidth]{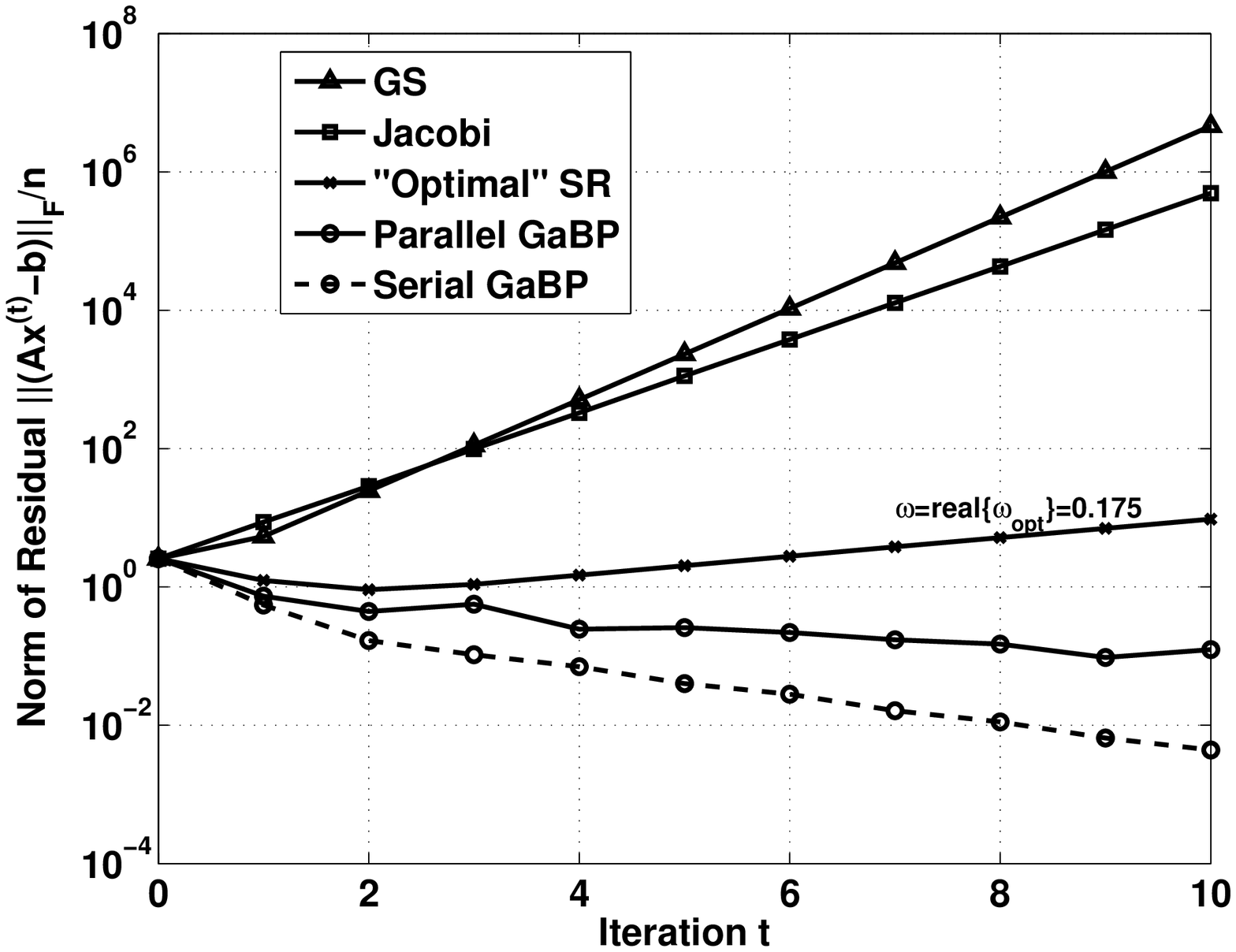}
 \label{fig_Asym}
\end{minipage}
\begin{minipage}[b]{0.5\linewidth}
\centering
   \includegraphics[width=\textwidth]{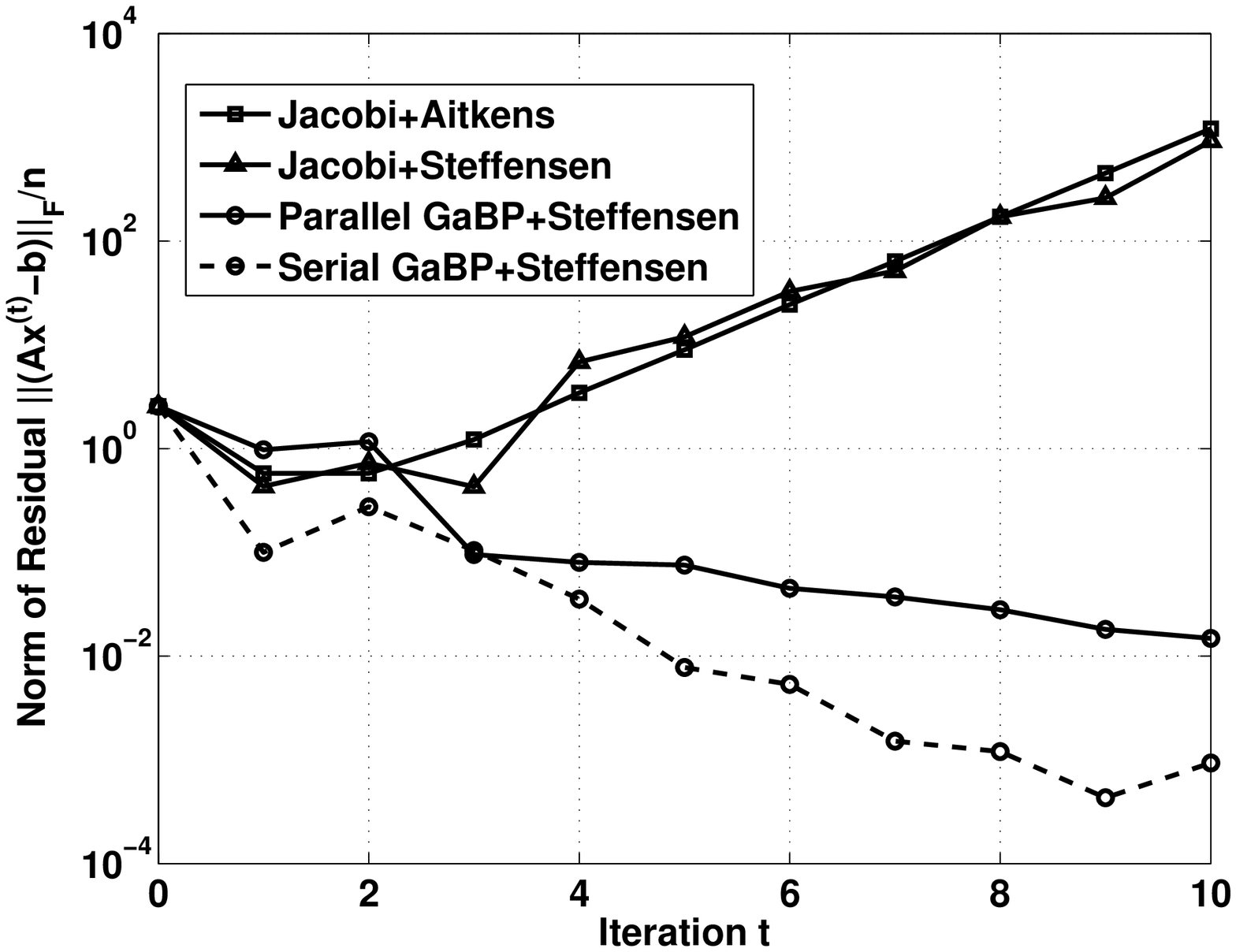}
 \label{fig_Asym_accel}
\end{minipage}
\caption{Convergence of an asymmetric $3 \times 3$ matrix. }
\end{figure}

\begin{figure}[h!]\label{fig_Asym_spiral}
\begin{center}
    \includegraphics[width=0.5\textwidth]{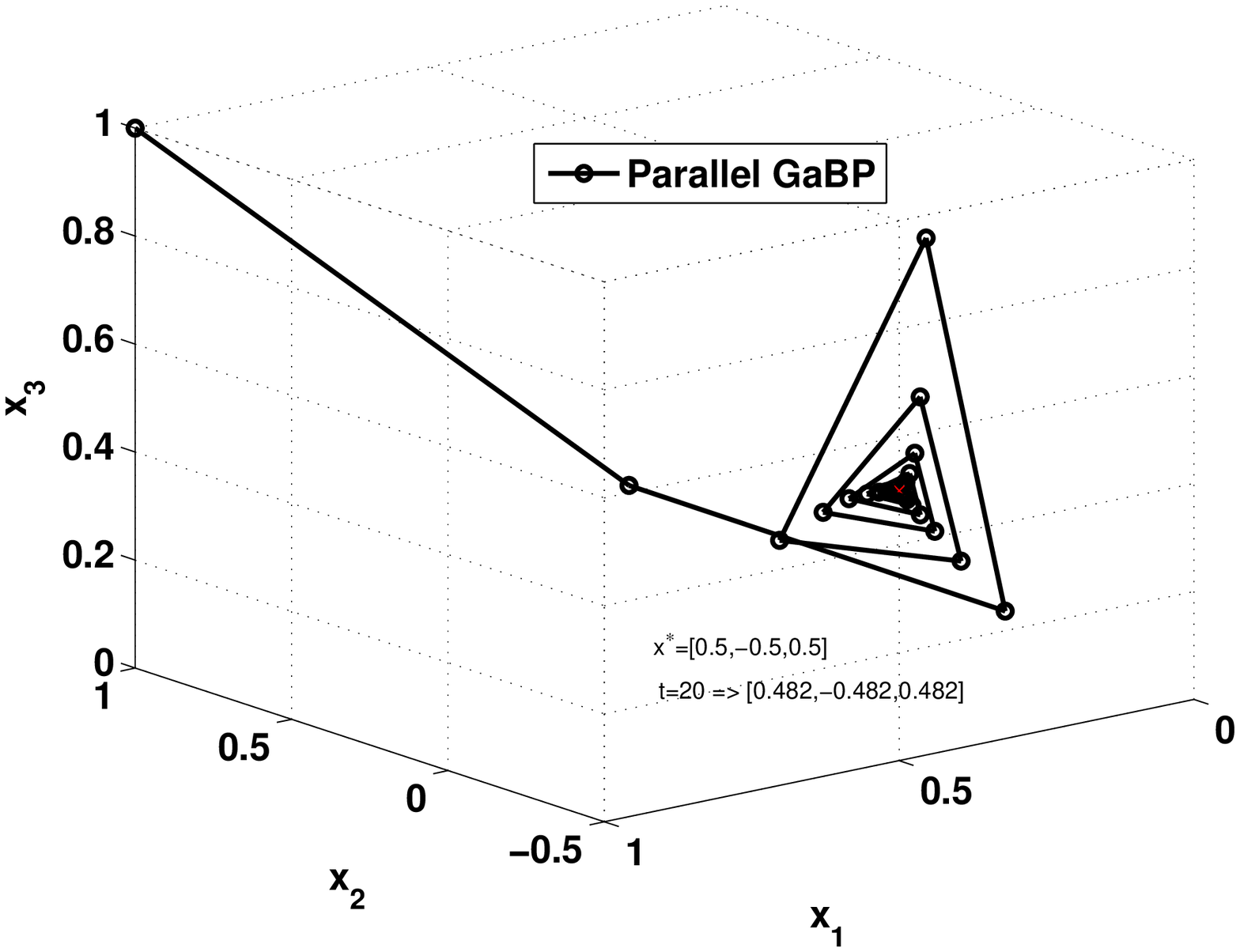}
  \caption{Convergence of a $3 \times 3$ asymmetric matrix, using 3D plot.}
\end{center}
\end{figure}


\section{Conclusion and Future Directions}
\label{sec:conclusion}
In this paper, we have established a powerful new connection
between the problem of solving a system of linear equations and
probabilistic inference on a suitable Gaussian graphical model.
Exploiting this connection, we have developed an iterative,
message-passing algorithm based upon Gaussian BP that can be used
as a low-complexity alternative to linear algebraic solutions
based upon direct matrix inversion. By its nature, the new
algorithm allows an efficient, distributed implementation.

There are numerous applications in mathematics and engineering to
which the new GaBP-based algorithm is applicable. We illustrated
its potential performance advantages in the context of linear
detection~\cite{Allerton,ISIT2}, but many other techniques
in digital communications requiring matrix inversion or
determinant computation, such as channel
precoding~\cite{BibDB:Precoding}, are amenable to the new
approach.
%
%

 The extension of this
technique to systems of linear equations over finite fields would
open up a wealth of other applications. For example, such a
development could lead to an iterative, message-passing algorithm
for efficient decoding of algebraic error-correcting codes, like
the widely-used class of BCH and Reed-Solomon codes.

\appendices
\section{Proof of Lemma~\ref{Lemma_Gaussian}}\label{App_Lemma}
\begin{proof}
Taking the product of the two Gaussian probability density functions \BE
    f_{1} (x)f_{2} (x)=\frac{\sqrt{P_{1}P_{2}}}{2\pi}\exp{\Big(-\big(P_{1}(x-\mu_{1})^{2}+P_{2}(x-\mu_{2})^{2}\big)/2\Big)}
\EE and completing the square, one gets \BE
    f_{1} (x)f_{2} (x)=\frac{C\sqrt{P}}{2\pi}\exp{\big(-P(x-\mu)^{2}/2\big)},
\EE with \BEA P&\triangleq&P_{1}+P_{2},\\
                \mu&\triangleq&P^{-1}(\mu_{1}P_{1}+\mu_{2}P_{2})
\EEA and the scalar constant determined by\BE
    C\triangleq\sqrt{\frac{P}{P_{1}P_{2}}}\exp{\Big(\big(P_{1}\mu_{1}^{2}(P^{-1}P_{1}-1)+P_{2}\mu_{2}^{2}(P^{-1}P_{2}-1)+2P^{-1}P_{1}P_{2}\mu_{1}\mu_{2}\big)/2\big)}.
\EE Hence, the product of the two Gaussian densities is $C\cdot\mathcal{N}(\mu,P^{-1})$.
\end{proof}

\section{Integrating over $x_{i}$}\label{App_Integral}
\begin{proof}
\BEA
    m_{ij}(x_j)&\propto&\int_{x_{i}} \psi_{ij}(x_i,x_j) \phi_{i}(x_i)
\prod_{k \in \textrm{N}(i)\setminus j} m_{ki}(x_i)
dx_{i}\\&\propto&\int_{x_{i}}\overbrace{\exp{(-x_{i}A_{ij}x_{j})}}^{\psi_{ij}(x_{i},x_{j})}\overbrace{\exp{(-P_{i\backslash
j}(x_{i}^{2}/2-\mu_{i\backslash j}x_{i}))}}^{\phi_{i}(x_i)
\prod_{k \in \textrm{N}(i) \backslash j}
m_{ki}(x_i)}dx_{i}\\&=&\int_{x_{i}}\exp{((-P_{i\backslash
j}x_{i}^{2}/2)+(P_{i\backslash j}\mu_{i\backslash
j}-A_{ij}x_{j})x_{i})}dx_{i}\\\label{eq_exponent}&\propto&\exp{((P_{i\backslash
j}\mu_{i\backslash j}-A_{ij}x_{j})^{2}/(2P_{i\backslash
j}))}\\&\propto&\mathcal{N}(\mu_{ij}=-P_{ij}^{-1}A_{ij}\mu_{i\backslash
j},P_{ij}^{-1} = -A_{ij}^{-2}P_{i \backslash
j}^{-1})\label{eq_IntegrationResult},\EEA where the
exponent~(\ref{eq_exponent}) is obtained by using the Gaussian
integral~(\ref{eq_GaussianIntegral}).
\end{proof}

\section{Maximizing over $x_{i}$}\label{App_Max}
\begin{proof}
\BEA
    m_{ij}(x_j)&\propto&\argmax_{x_{i}} \psi_{ij}(x_i,x_j) \phi_{i}(x_i)
\prod_{k \in \textrm{N}(i)\setminus j} m_{ki}(x_i)\\&\propto&\argmax_{x_{i}}\overbrace{\exp{(-x_{i}A_{ij}x_{j})}}^{\psi_{ij}(x_{i},x_{j})}\overbrace{\exp{(-P_{i\backslash
j}(x_{i}^{2}/2-\mu_{i\backslash j}x_{i}))}}^{\phi_{i}(x_i)
\prod_{k \in \textrm{N}(i) \backslash j}
m_{ki}(x_i)}\\&=&\argmax_{x_{i}}\exp{((-P_{i\backslash
j}x_{i}^{2}/2)+(P_{i\backslash j}\mu_{i\backslash
j}-A_{ij}x_{j})x_{i})}. \EEA
Hence, $x_{i}^{\textrm{max}}$, the value of $x_{i}$ maximizing the product $\psi_{ij}(x_i,x_j) \phi_{i}(x_i)
\prod_{k \in \textrm{N}(i)\setminus j} m_{ki}(x_i)$ is given by equating its derivative w.r.t. $x_{i}$ to zero, yielding \BE
x_{i}^{\textrm{max}}=\frac{P_{i\backslash j}\mu_{i\backslash
j}-A_{ij}x_{j}}{P_{i\backslash j}}.
\EE
Substituting $x_{i}^{\textrm{max}}$ back into the product, we get \BEA
m_{ij}(x_j)&\propto&\exp{((P_{i\backslash
j}\mu_{i\backslash j}-A_{ij}x_{j})^{2}/(2P_{i\backslash
j}))}\\&\propto&\mathcal{N}(\mu_{ij}=-P_{ij}^{-1}A_{ij}\mu_{i\backslash
j},P_{ij}^{-1} = -A_{ij}^{-2}P_{i \backslash j}),\EEA which is identical to the result obtained when eliminating
$x_{i}$ via integration~(\ref{eq_IntegrationResult}).
\end{proof}

\section{Quadratic Min-Sum Message Passing algorithm}
\label{MinSum}
The quadratic Min-Sum message passing algorithm was initially presented in \cite{MinSum}.
It is a variant of the max-product algorithm, with underlying Gaussian distributions.
The quadratic Min-Sum algorithm is an iterative algorithm for solving a quadratic cost function. Not surprisingly, as we have shown in Section \ref{MaxProductRule} that the Max-Product and the Sum-Product algorithms are identical when the underlying distributions are Gaussians. In this contribution, we show that the quadratic Min-Sum algorithm
is identical to the GaBP algorithm, although it was was derived differently.

In \cite{MinSum} the authors discuss the application for solving linear system of equations
using the Min-Sum algorithm. Our work \cite{Allerton} was done in parallel to their work, where
both papers appeared in the 45th Allerton conference.

\begin{thm}
The Quadratic Min-Sum algorithm is an instance of the GaBP algorithm.
\end{thm}
\begin{proof}[\bf Proof]
We start in the quadratic parameter updates:
\[ \gamma_{ij} = \frac{1}{1 - \Sigma_{u \in N(i) \backslash j}\Gamma^2_{ui} \gamma_{ui}} =
\overbrace{(\overbrace{1}^{A_{ii}} - \Sigma_{u \in N(i) \backslash j}\overbrace{\Gamma_{ui}}^{A_{ui}} \overbrace{\gamma_{ui}}^{P_{ui}^{-1}} \overbrace{\Gamma_{iu}}^{A_{iu}})^{-1}}^{P_{i \backslash j}^{-1}}
\]
Which is equivalent to \ref{eq_prec}.
Regarding the mean parameters,
\[ z_{ij} = \frac{\Gamma_{ij} }{1 - \Sigma_{u \in N(i) \backslash j}\Gamma^2_{ui} \gamma_{ui} } (h_i - \Sigma_{u \in N(i) \backslash j} z_{ui}) = \overbrace{ \overbrace{\Gamma_{ij}}^{A_{ij}} \overbrace{\gamma_{ij}}^{(P_{i \backslash j})^{-1}} (\overbrace{h_i}^{b_i} - \Sigma_{u \in N(i) \backslash j} z_{ui})}^{\mu_{i \backslash j}} \]
Which is equivalent to \ref{eq_mean}.
\end{proof}

For simplicity of notations, we list the different notations in
Min-Sum paper vs. our notations:
\begin{table}[h!]
\caption{Notations of Min-Sum \cite{MinSum} vs. GaBP}
\begin{center}
\begin{tabular}{|c|c|c|}
  \hline
  Min-Sum~\cite{MinSum} & GaBP~\cite{Allerton} & comments\\
  \hline
  $\gamma_{ij}^{(t+1)}$ & $P_{i \backslash j}^{-1} $ & quadratic parameters / product rule precision from $i$ to $j$ \\
  $z_{ij}^{(t+1)}$ & $\mu_{i \backslash j}$ & linear parameters / product rule mean rom $i$ to $j$ \\
  $h_i$ & $b_i$ & prior mean of node $i$\\
  $A_{ii}$ & $1$ & prior precision of node $i$\\
  $x_i$ & $x_i$ & posterior mean of node $i$ \\
  $-$ & $P_{i}$ & posterior precision of node $i$ \\
  $\Gamma_{ij}$ & $A_{ij}$ & covariance of nodes $i$ and $j$ \\
  \hline
\end{tabular}
\end{center}
\end{table}
As shown above, the Min-Sum algorithm assumes the covariance matrix $\Gamma$ is first normalized s.t. the main
diagonal entries (the variances) are all one.
The messages sent in the Min-Sum algorithm are called linear parameters (which are equivalent to the mean messages in GaBP) and quadratic parameters (which are equivalent to variances). The difference between the algorithm is that in the GaBP algorithm, a node computes the product rule and the integral, and sends the result to its neighbor. In the Min-Sum algorithm, a node computes the product rule, sends the intermediate result, and
the receiving node computes the integral. In other words, the same computation is performed but on different
locations. In the Min-Sum algorithm terminology, the messages are linear and quadratic parameters vs.
Gaussians in our terminology.

\bibliographystyle{IEEEtran}   
\footnotesize
\bibliography{IEEEabrv,GaBP_J_v8}       

\begin{thebibliography}{10}
\providecommand{\url}[1]{#1}
\csname url@samestyle\endcsname
\providecommand{\newblock}{\relax}
\providecommand{\bibinfo}[2]{#2}
\providecommand{\BIBentrySTDinterwordspacing}{\spaceskip=0pt\relax}
\providecommand{\BIBentryALTinterwordstretchfactor}{4}
\providecommand{\BIBentryALTinterwordspacing}{\spaceskip=\fontdimen2\font plus
\BIBentryALTinterwordstretchfactor\fontdimen3\font minus
  \fontdimen4\font\relax}
\providecommand{\BIBforeignlanguage}[2]{{%
\expandafter\ifx\csname l@#1\endcsname\relax
\typeout{** WARNING: IEEEtran.bst: No hyphenation pattern has been}%
\typeout{** loaded for the language `#1'. Using the pattern for}%
\typeout{** the default language instead.}%
\else
\language=\csname l@#1\endcsname
\fi
#2}}
\providecommand{\BIBdecl}{\relax}
\BIBdecl

\bibitem{BibDB:BookMatrix}
G.~H. Golub and C.~F.~V. Loan, Eds., \emph{Matrix Computation}, 3rd~ed.\hskip
  1em plus 0.5em minus 0.4em\relax The Johns Hopkins University Press, 1996.

\bibitem{BibDB:BookAxelsson}
O.~Axelsson, \emph{Iterative Solution Methods}.\hskip 1em plus 0.5em minus
  0.4em\relax Cambridge, UK: Cambridge University Press, 1994.

\bibitem{BibDB:BookSaad}
Y.~Saad, Ed., \emph{Iterative methods for Sparse Linear Systems}.\hskip 1em
  plus 0.5em minus 0.4em\relax PWS Publishing company, 1996.

\bibitem{BibDB:BookPearl}
J.~Pearl, \emph{Probabilistic Reasoning in Intelligent Systems: Networks of
  Plausible Inference}.\hskip 1em plus 0.5em minus 0.4em\relax San Francisco:
  Morgan Kaufmann, 1988.

\bibitem{BibDB:BookJordan}
M.~I. Jordan, Ed., \emph{Learning in Graphical Models}.\hskip 1em plus 0.5em
  minus 0.4em\relax Cambridge, MA: The MIT Press, 1999.

\bibitem{MinSum}
C.~Moallemi and B.~V. Roy, ``Convergence of the min-sum algorithm for convex
  optimization,'' in \emph{Proc. of the 45th Allerton Conference on
  Communication, Control and Computing}, Monticello, IL, September 2007.

\bibitem{Allerton}
D.~Bickson, O.~Shental, P.~H. Siegel, J.~K. Wolf, and D.~Dolev, ``Linear
  detection via belief propagation,'' in \emph{Proc. 45th Allerton Conf. on
  Communications, Control and Computing}, Monticello, IL, USA, Sep. 2007.

\bibitem{BibDB:Weiss01Correctness}
Y.~Weiss and W.~T. Freeman, ``Correctness of belief propagation in {Gaussian}
  graphical models of arbitrary topology,'' \emph{Neural Computation}, vol.~13,
  no.~10, pp. 2173--2200, 2001.

\bibitem{BibDB:jmw_walksum_nips}
J.~K. Johnson, D.~M. Malioutov, and A.~S. Willsky, ``Walk-sum interpretation
  and analysis of {Gaussian} belief propagation,'' in \emph{Advances in Neural
  Information Processing Systems 18}, Y.~Weiss, B.~Sch\"{o}lkopf, and J.~Platt,
  Eds.\hskip 1em plus 0.5em minus 0.4em\relax Cambridge, MA: MIT Press, 2006,
  pp. 579--586.

\bibitem{BibDB:mjw_walksum_jmlr06}
D.~M. Malioutov, J.~K. Johnson, and A.~S. Willsky, ``Walk-sums and belief
  propagation in {Gaussian} graphical models,'' \emph{Journal of Machine
  Learning Research}, vol.~7, Oct. 2006.

\bibitem{ISIT1}
O.~Shental, P.~H.~S. D.~Bickson, J.~K. Wolf, and D.~Dolev, ``Gaussian belief
  propagation solver for systems of linear equations,'' in \emph{IEEE Int.
  Symp. on Inform. Theory (ISIT)}, Toronto, Canada, July 2008.

\bibitem{ISIT2}
D.~Bickson, O.~Shental, P.~H. Siegel, J.~K. Wolf, and D.~Dolev, ``Gaussian
  belief propagation based multiuser detection,'' in \emph{IEEE Int. Symp. on
  Inform. Theory (ISIT)}, Toronto, Canada, July 2008.

\bibitem{grant99iterative}
A.~Grant and C.~Schlegel, ``Iterative implementations for linear multiuser
  detectors,'' \emph{{IEEE} Trans. Commun.}, vol.~49, no.~10, pp. 1824--1834,
  Oct. 2001.

\bibitem{BibDB:TanRasmussen}
P.~H. Tan and L.~K. Rasmussen, ``Linear interference cancellation in {CDMA}
  based on iterative techniques for linear equation systems,'' \emph{{IEEE}
  Trans. Commun.}, vol.~48, no.~12, pp. 2099--2108, Dec. 2000.

\bibitem{BibDB:YenerEtAl}
A.~Yener, R.~D. Yates, , and S.~Ulukus, ``{CDMA} multiuser detection: {A}
  nonlinear programming approach,'' \emph{{IEEE} Trans. Commun.}, vol.~50,
  no.~6, pp. 1016--1024, Jun. 2002.

\bibitem{BibDB:BookHenrici}
P.~Henrici, \emph{Elements of Numerical Analysis}.\hskip 1em plus 0.5em minus
  0.4em\relax John Wiley and Sons, 1964.

\bibitem{BibDB:AjiMcEliece}
S.~M. Aji and R.~J. McEliece, ``The generalized distributive law,''
  \emph{{IEEE} Trans. Inf. Theory}, vol.~46, no.~2, pp. 325--343, Mar. 2000.

\bibitem{BibDB:FactorGraph}
F.~Kschischang, B.~Frey, and H.~A. Loeliger, ``Factor graphs and the
  sum-product algorithm,'' \emph{{IEEE} Trans. Inf. Theory}, vol.~47, pp.
  498--519, Feb. 2001.

\bibitem{BibDB:ElidanEtAl}
G.~Elidan, Mcgraw, and D.~Koller, ``Residual belief propagation: Informed
  scheduling for asynchronous message passing,'' July 2006.

\bibitem{Max-product}
Y.~Weiss and W.~T. Freeman, ``On the optimality of solutions of the max-product
  belief-propagation algorithm in arbitrary graphs,'' in \emph{Information
  Theory, IEEE Transactions on}, vol.~47, no.~2, 2001, pp. 736--744.

\bibitem{BibDB:BCJR}
L.~R. Bahl, J.~Cocke, F.~Jelinek, and J.~Raviv, ``Optimal decoding of linear
  codes for minimizing symbol error rate,'' \emph{{IEEE} Trans. Inf. Theory},
  vol.~20, no.~3, pp. 284--287, Mar. 1974.

\bibitem{Viterbi}
A.~Viterbi, ``Error bounds for convolutional codes and an asymptotically
  optimum decoding algorithm,'' in \emph{Information Theory, IEEE Transactions
  on}, vol.~13, no.~2, 1967, pp. 260--269.

\bibitem{BibDB:BookBertsekasTsitsiklis}
D.~P. Bertsekas and J.~N. Tsitsiklis, \emph{Parallel and Distributed
  Calculation. Numerical Methods.}\hskip 1em plus 0.5em minus 0.4em\relax
  Prentice Hall, 1989.

\bibitem{Damping}
K.~M. Murphy, Y.~Weiss, and M.~I. Jordan, ``Loopy belief propagation for
  approximate inference: {An} empirical study,'' in \emph{Proc. of {UAI}},
  1999.

\bibitem{BibDB:BookVerdu}
S.~Verd\'{u}, \emph{Multiuser Detection}.\hskip 1em plus 0.5em minus
  0.4em\relax Cambridge, UK: Cambridge University Press, 1998.

\bibitem{BibDB:BookProakis}
J.~G. Proakis, \emph{Digital Communications}, 4th~ed.\hskip 1em plus 0.5em
  minus 0.4em\relax New York, USA: McGraw-Hill, 2000.

\bibitem{BibDB:Kabashima}
Y.~Kabashima, ``A {CDMA} multiuser detection algorithm on the basis of belief
  propagation,'' \emph{J. Phys. A: Math. Gen.}, vol.~36, pp. 11\,111--11\,121,
  Oct. 2003.

\bibitem{BibDB:ShentalITW}
O.~Shental, N.~Shental, A.~J. Weiss, and Y.~Weiss, ``Generalized belief
  propagation receiver for near-optimal detection of two-dimensional channels
  with memory,'' in \emph{Proc. {IEEE} Information Theory Workshop ({ITW})},
  San Antonio, Texas, USA, Oct. 2004.

\bibitem{BibDB:TanakaOkada}
T.~Tanaka and M.~Okada, ``Approximate belief propagation, density evolution,
  and statistical neurodynamics for {CDMA} multiuser detection,'' \emph{{IEEE}
  Trans. Inf. Theory}, vol.~51, no.~2, pp. 700--706, Feb. 2005.

\bibitem{BibDB:MontanariTse}
A.~Montanari and D.~Tse, ``Analysis of belief propagation for non-linear
  problems: The example of {CDMA} (or: How to prove {Tanaka's} formula),'' in
  \emph{Proc. {IEEE} Inform. Theory Workshop ({ITW})}, {Punta del Este},
  Uruguay, Mar. 2006.

\bibitem{BibDB:ConfWangGuo}
C.~C. Wang and D.~Guo, ``Belief propagation is asymptotically equivalent to
  {MAP} detection for sparse linear systems,'' in \emph{Proc. 44th Allerton
  Conf. on Communications, Control and Computing}, Monticello, IL, USA, Sep.
  2006.

\bibitem{LDF05}
C.~Leibig, A.~Dekorsy, and J.~Fliege, ``Power control using {Steffensen}
  iterations for {CDMA} systems with beamforming or multiuser detection,'' in
  \emph{Proc. {IEEE} International Conference on Communications ({ICC})},
  Seoul, Korea, 2005.

\bibitem{ECCS08}
D.~Bickson, D.~Dolev, and E.~Yom-Tov, ``A gaussian belief propagation solver
  for large scale support vector machines,'' in \emph{5th European Conference
  on Complex Systems}, Jerusalem, Sept. 2008.

\bibitem{Rating}
D.~Bickson, D.~Malkhi, and L.~Zhou, ``{Peer-to-Peer} rating,'' in \emph{7th
  IEEE {P2P} computing}, Galway, Ireland, 2007.

\bibitem{PPNA08}
D.~Bickson and D.~Malkhi, ``A unifying framework for rating users and data
  items in peer-to-peer and social networks,'' in \emph{Peer-to-Peer Networking
  and Applications (PPNA) Journal, Springer-Verlag}, April 2008.

\bibitem{BibDB:MontanariEtAl}
A.~Montanari, B.~Prabhakar, and D.~Tse, ``Belief propagation based multi-user
  detection,'' in \emph{Proc. 43th Allerton Conf. on Communications, Control
  and Computing}, Monticello, IL, USA, Sep. 2005.

\bibitem{BibDB:Precoding}
B.~R. Voj\v{c}i\'{c} and W.~M. Jang, ``Transmitter precoding in synchronous
  multiuser communications,'' \emph{{IEEE} Trans. Commun.}, vol.~46, no.~10,
  pp. 1346--1355, Oct. 1998.

\end{thebibliography}
\newpage

\end{document}